\documentclass[a4paper,11pt]{article}
\usepackage{jheppub} 
\usepackage{lineno}
\usepackage[dvipsnames]{xcolor}
\usepackage[normalem]{ulem}

\usepackage{subcaption}

\newcommand{\be}{\begin{equation}}
\newcommand{\ee}{\end{equation}}
\newcommand{\bea}{\begin{eqnarray}}
\newcommand{\eea}{\end{eqnarray}}
\def\vec{\mathbf}
\def\veck{\vec{k}}
\def\vecx{\vec{x}}

\title{\boldmath Double-well instantons in finite volume}

\author[a]{Wen-Yuan Ai,}
\author[a]{Jean Alexandre,}
\author[b,1]{Matthias Carosi,\note{Corresponding author}}
\author[b]{Björn Garbrecht,}
\author[a]{and Silvia Pla}
\affiliation[a]{Department of Physics, King's College London,\\ Strand, London  WC2RS 2LS, United Kingdom}
\affiliation[b]{Physik-Department, Technische Universit\"at M\"unchen,\\ James-Franck-Str., 85748 Garching, Germany}

\emailAdd{wenyuan.ai@kcl.ac.uk}
\emailAdd{jean.alexandre@kcl.ac.uk}
\emailAdd{matthias.carosi@tum.de}
\emailAdd{garbrecht@tum.de}\emailAdd{silvia.pla\_garcia@kcl.ac.uk}

\abstract{Assuming a toroidal space with finite volume, we derive analytically the full one-loop vacuum energy for a scalar field tunnelling between two degenerate vacua, taking into account discrete momentum. The Casimir energy is computed for an arbitrary number of dimensions using the Abel-Plana formula, while the one-loop instanton functional determinant is evaluated using the Green's functions for the fluctuation operators. The resulting energetic properties are non-trivial: both the Casimir effect and tunnelling contribute to the Null Energy Condition violation, arising from a non-extensive true vacuum energy.
We discuss the relevance of this mechanism to induce a cosmic bounce, requiring no modified gravity or exotic matter.
}

\begin{document}
\begin{flushright}
\raggedleft
    \small KCL-PH-TH/2024-04 \\ TUM-HEP-1497/24
\end{flushright}
\maketitle

\section{Introduction}
\label{sec: intro}

Quantum Field Theory (QFT) usually assumes an infinite volume, in which case the energy levels are continuous, 
and tunnelling between degenerate vacua is completely suppressed. 
In finite volume things are different though: energies are quantised, and tunnelling between degenerate vacua is allowed, as in Quantum Mechanics. 
The first feature leads to the known Casimir effect (see~\cite{Bordag} for a review) while the second is at the origin of convexity of the effective potential~\cite{Symanzik:1969ek,Coleman:1974jh,Iliopoulos:1974ur,Haymaker:1983xk,Fujimoto:1982tc,Bender:1983nc,Hindmarsh:1985nc,Alexandre:2012ht,Plascencia:2015pga,Millington:2019nkw} and therefore restoration of a symmetric true vacuum.
In the present article, we consider the energetic consequences of the interplay between the above finite volume effects, 
that we compare in the specific situation where space has the topology of a three-torus.

The Casimir energy of a system in a finite volume is defined as the difference between the ground state energy calculated with 
discrete and continuous momenta. It is known that the result is highly dependent on the geometry of the confining space, as well as the 
boundary conditions satisfied by the field inside this space. 
For a field with mass $m$, the Casimir effect is usually suppressed as $\approx\exp(-mL)$ for $mL\gg 1$, where $L$ is a typical length of the system. Tunnelling, on the other hand, is typically suppressed as the exponential of the volume $\approx\exp[-(mL)^3]$, where $(mL)^3$ is the action of a 
Euclidean-time-dependent and spatially homogeneous instanton connecting the two degenerate 
vacua.\footnote{For this reason tunnelling between degenerate vacua is completely suppressed in the limit of infinite volume, where 
spontaneous symmetry breaking occurs instead.} As a result, tunnelling is negligible compared to the Casimir effect if we consider the 
situation $mL\gg1$.\footnote{This is the case of three space dimensions, and of course things could be different in one space dimension.} 

In the situation where $mL\lesssim1$, tunnelling and Casimir effects could be of the same order, potentially leading to new phenomena. 
The effective theory taking into account tunnelling of a real scalar field between two degenerate vacua has been studied in~\cite{Alexandre:2022qxc,Alexandre:2023iig}.
It was found that the effective potential is indeed convex, with a symmetric true vacuum at $\left<\phi\right>=0$ and a corresponding ground state energy lower than the original degenerate vacua, similarly to what happens in Quantum Mechanics.
This result was obtained by neglecting the spatially inhomogeneous fluctuations around the instanton configuration, as well as the Casimir effect.
In the present work, we refine this analysis by including all quantum fluctuations and using a full analytic result for the Casimir energy.

We note here that $O(4)$-symmetric saddle points, representing bubbles of true vacuum developing in a false vacuum, are allowed in the situation of non-degenerate vacua only~\cite{Coleman:1977py,Callan:1977pt} (see~\cite{Andreassen:2016cvx,Ai:2019fri} for reviews).
Degenerate vacua would lead to bubbles of infinite radius, which justifies only looking at homogeneous saddle points in the present work.
We also note that the tunnelling-induced symmetry restoration described here happens at zero temperature and is unrelated to high-temperature symmetry restoration. 

An important feature related to the above finite-volume effects is the violation of the Null Energy Condition 
(NEC - see~\cite{Rubakov:2014jja,Kontou:2020bta} for reviews), which is known in the case of the Casimir effect, 
and which has been shown in~\cite{Alexandre:2022qxc,Alexandre:2023iig} for tunnelling between degenerate vacua.
NEC violation in both cases is related to a non-trivial dependence on the typical length $L$ of the system considered,
which implies a non-extensive ground state energy and leads to remarkable energetic properties.
In this article we derive these effects precisely, taking into account the full one-loop quantum fluctuations about the instanton, with discrete momenta. \\

Section~\ref{sec: the model} sets the framework, where we define the semiclassical approximation for the partition function of 
a self-interacting scalar field with a double-well potential.
We consider the same instanton dilute gas as the one used in Quantum Mechanics, based on
homogeneous and Euclidean-time-dependent configurations \cite{Kleinert:2004ev}.

In Section~\ref{sec: casimir energy} we derive the Casimir energy for a massive scalar field in a three-torus, using the Abel-Plana formula. 
This is done first in one space dimension and then in arbitrary dimensions, based on a recursion formula which is derived in Appendix~\ref{app: casimir energy technical details}, 
together with the details of the calculation for two and three space dimensions.

The tunnelling contribution is then described in Section~\ref{sec: tunnelling}, where the full one-loop functional determinant for fluctuations about the instanton is obtained based on the method of the Green's function described therein. Throughout the calculation, the discrete nature of the momenta is kept, and no further approximation is made.
This calculation is done for the first time, and all the details are therefore given in this section. 
The true ground state energy is then discussed, and we find that tunnelling can be of the same order of magnitude as the Casimir effect for $mL\lesssim1$.

In Section~\ref{sec: applications} we discuss the possibility of inducing a cosmic bounce from the NEC violation mechanism obtained here,
without modified gravity or exotic matter, as was already described in~\cite{Alexandre:2023pkk,Alexandre:2023bih}, 
but where quantum fluctuations were calculated with continuous momenta.

\section{True ground state energy}\label{sec: the model}

We consider a real scalar field in a symmetric double-well potential and in finite volume. The Quantum Mechanical analogy, corresponding to field theory in zero spatial dimensions, is well-known from Ref.~\cite{Coleman:1985rnk}.
Here, we choose space to be a three-torus with size $L$ in the three directions, thus corresponding to the volume $L^3$.
The Euclidean action is 
\begin{equation}\label{eqn: euclidean action}
    S_E = \int_{-\beta/2}^{\beta/2} d\tau \int d^3\vecx \left( \frac{1}{2} (\partial_\tau \phi)^2 + \frac{1}{2} (\nabla \phi)^2 +U(\phi) \right) ~,
\end{equation}
where
\begin{align}
    U(\phi)=\frac{\lambda}{4!} (\phi^2-v^2)^2 +U_0\,,
\end{align}
and $\tau = it$ is the imaginary time. We will consider the zero-temperature limit $\beta\to\infty$. 
The energy density $U_0$, which fixes the origin of energies, corresponds to a cosmological constant in the discussion on Cosmology.

\begin{figure}
    \centering
         \includegraphics[width=0.7\textwidth]{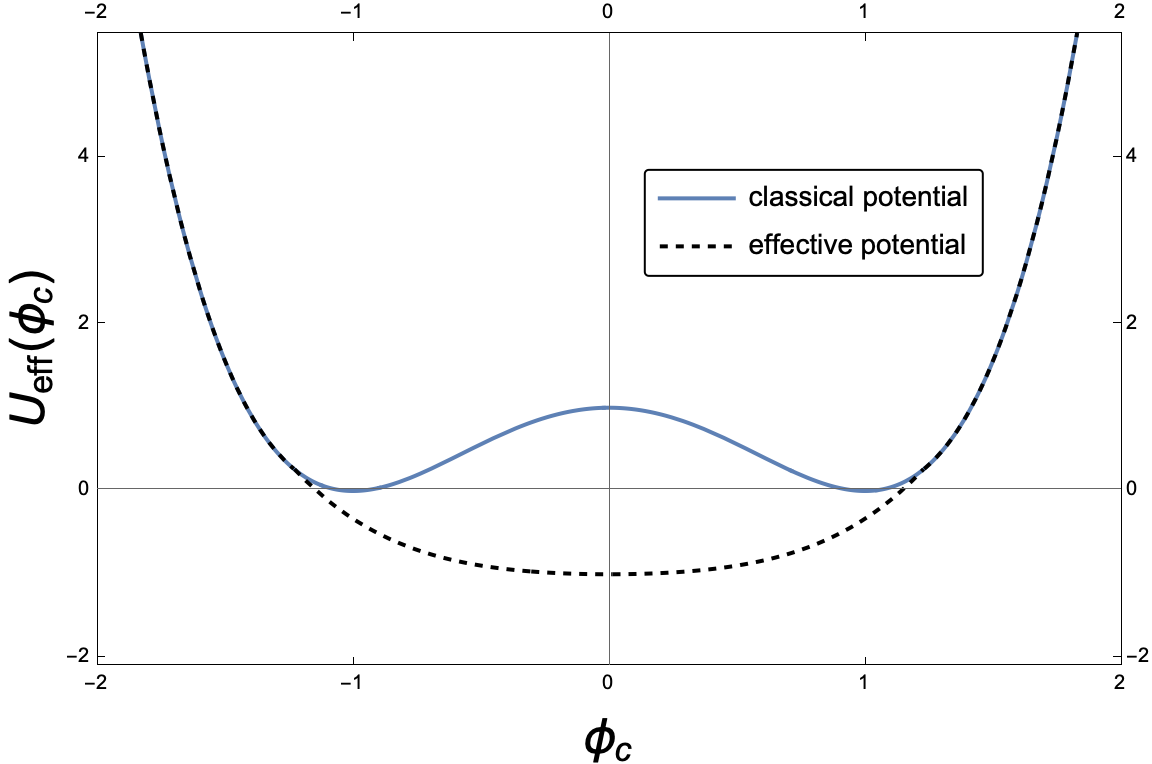}  
    \caption{{\small The classical potential (solid line) and the effective potential (dashed line) in a finite volume. Tunnelling between the degenerate minima in the classical potential restores the symmetry, giving a symmetric true vacuum at $\phi_c=0$.}}
    \label{fig: effective potential}
\end{figure}

As discussed in the Introduction, in this context the scalar field can tunnel between the two degenerate bare minima $\pm v$, 
resulting in an effective theory with a symmetric vacuum state located at $\left<\phi\right>=0$ (see Figure~\ref{fig: effective potential}). 
The explicit effective potential has been calculated at quadratic order in~\cite{Alexandre:2022qxc,Alexandre:2023iig}, 
where one finds a positive mass term, therefore showing convexity. 
In the present article, we focus on the true vacuum expectation value (vev) from the beginning and thus we consider a vanishing 
source. 
Indeed, finite volume implies a one-to-one mapping between the source and the vev, and thus a unique Legendre transform 
to obtain the one-particle-irreducible effective action. In this mapping a vanishing source corresponds to a vanishing vev.

\subsection{Semi-classical approximation}

We focus here on the fundamental quantity, which is the true ground state energy
\begin{equation}
\label{eqn: E0 from Z}
    E_0 = - \lim_{\beta\to\infty} \frac{1}{\beta} \log Z_E = - \lim_{\beta\to\infty} \frac{1}{\beta} \log \int_{\phi\left(-\frac{\beta}{2},\vecx\right)=\phi\left(\frac{\beta}{2},\vecx\right)} \mathcal{D}[\phi]~e^{-S_E[\phi]} ~,
\end{equation}
$\beta$ being  the inverse temperature, $Z_E$ the Euclidean partition function and $S_E$ the Euclidean action~\eqref{eqn: euclidean action}. Above, we have indicated the periodic boundary conditions in Euclidean time when evaluating the canonical thermal partition function.
We will evaluate $Z_E$ in the context of the semi-classical approximation, where the contribution of each dominant configuration (the saddle points)
is independent from the others, such that 
\begin{equation}
\begin{aligned}
    Z_E = & \int \mathcal{D}[\phi]~ e^{-S_E[\phi]} \approx \sum_{\overline \phi \: \mathrm{saddle}} [\overline\phi]~,
\end{aligned}
\end{equation}
where $[\overline{\phi}]$ is the one-loop approximation of the individual path-integral corresponding to a unique saddle point $\overline{\phi}$
\begin{align}\label{eqn: saddle point approximation}
    \relax[\overline{\phi}] & =  \int \mathcal{D}[\eta]~ e^{-S_E[\overline{\phi} + \eta]} \notag \\
    &\approx  e^{-S[\overline{\phi}]} \int \mathcal{D}[\eta]~ e^{-\frac{1}{2}\int \eta S_E''[\overline{\phi}] \eta} \: = \:  e^{-S_E[\overline{\phi}]} \left( \det S_E''[\overline{\phi}] \right)^{-\frac{1}{2}} ~,
\end{align}
where $S''_E[\Bar{\phi}]$ denotes the fluctuation operator about the saddle point $\bar\phi$, and we have used the simplified notation for the integral in the exponent.
Equation~\eqref{eqn: saddle point approximation} requires further treatment in case the saddle point $\overline{\phi}$ explicitly breaks some symmetry of the action. In this case, the operator $S_E''[\overline{\phi}]$ has one or more zero modes, over which the integral is not Gaussian. The integral over the zero modes, however, can be traded for an integral over the collective coordinates of the saddle point.

The homogeneous saddle points $\overline{\phi}$ are the solutions to the classical (Euclidean) equations of motion
\begin{equation}\label{eqn: eom}
    \Ddot{\phi} + \frac{\lambda v^2}{6} \phi - \frac{\lambda}{6}\phi^3 = 0\,,
\end{equation}
where the dot stands for the derivative with respect to the Euclidean time, $\Ddot{\phi} = \frac{\partial^2\phi}{\partial\tau^2}$.

In order for the saddle point to have a finite Euclidean action for $\beta\rightarrow\infty$, it must approach asymptotically either of the two minima of the classical potential $\pm v$ at $\tau=\pm\frac{\beta}{2}$. This simplifies the analysis since we only need to look for saddle points satisfying the boundary conditions $\phi\left(-\frac{\beta}{2}\right)=\phi\left(\frac{\beta}{2}\right)=\pm v$.
There are two trivial solutions in this model, $\phi\equiv \pm v$. The contribution of these two configuration reads
\begin{equation}\label{eqn:constant contribution}
    [+v] = [-v] = e^{-L^3 \beta U_0} \left( \det S_E''[v] \right)^{-\frac{1}{2}} = e^{-L^3 \beta U_0} \left( \det (-\partial_\tau^2 - \nabla^2 + m^2) \right)^{-\frac{1}{2}}\,,
\end{equation}
where $m^2\equiv U''(\phi)|_{\phi=\pm v} = \frac{\lambda v^2}{3}$. We postpone the calculation of the functional determinant to Section~\ref{sec: casimir energy}.

\subsection{Instanton gas}
\label{sec: gas}

Equation~\eqref{eqn: eom} also admits Euclidean-time-dependent but spatially homogeneous solutions, the instanton and the anti-instanton, i.e., the kink ($K$) and the anti-kink ($\overline{K}$) in Euclidean time
\begin{equation}\label{eqn: kink}
    \phi_{K/\overline{K}}(\tau) = \pm v \tanh \omega (\tau - \tau_1) \,, \qquad \mathrm{where} \qquad \omega^2 = \frac{\lambda v^2}{12} = \frac{m^2}{4}\,,
\end{equation}
and $\tau_1$ is the location of the (anti-)kink.
Strictly speaking, these saddle points do not contribute to the canonical partition function because they are not periodic in $\tau$. However, some combinations of them called multi-instantons do, and their contribution can be written in terms of single kink and anti-kink contributions as we will describe below. The contribution of a single (anti-)kink reads
\begin{equation}\label{eqn:kink contribution}
    \left[K\right] = \left[\overline{K}\right] = \int_{-\beta/2}^{\beta/2} d\tau_1 [I][v] = \beta [I][v]\,,
\end{equation}
where we isolated the contribution of the centre of the (anti-)instanton (the ``jump")
\begin{align}\label{eqn:jump contribution}
    \relax[I] = &\: \sqrt{\frac{K}{2\pi}} e^{-K} \left(\frac{\sideset{}{'}\det S_E''[\phi_K]}{\det S_E''[v]} \right)^{-\frac{1}{2}} \notag \\
    = & \:\sqrt{\frac{K}{2\pi}} e^{-K} \left(\frac{\sideset{}{'}\det \left(-\partial_\tau^2 - \nabla^2 + \frac{m^2}{2} \left(\frac{3\phi_K^2}{v^2} - 1\right)\right)}{\det (-\partial_\tau^2 - \nabla^2 + m^2)} \right)^{-\frac{1}{2}} ~,
\end{align}
and $\det^\prime$ is the fluctuation determinant without the contribution of the zero mode. 
The latter arises from the invariance of the system under translation of the jump and leads to the integration over the variable $\tau_1$ in Eq.~\eqref{eqn:kink contribution}. We exchange the integration of fluctuations in the direction of the time-translational zero mode $\sim\partial_\tau\phi_K$ for an integration over the location of the jump. In fact, the jump can be located at any point in the time interval $(-\beta/2,\beta/2)$, and we must sum over all possible such positions. This leads to the overall factor of $\beta$.
The factor of $\sqrt{\frac{K}{2\pi}}$ is the Jacobian of the transformation to the collective coordinate $\tau_1$.
Finally, $K$ is the classical (Euclidean) action of the kink configuration, without $U_0$, 
\begin{equation}
    K = \frac{2(m L)^3}{\lambda}~.
\end{equation}

We now turn to the contribution of the multi-instantons, together with their zero modes.  
These correspond to back-and-forth oscillations between the two maxima of the upside-down potential. 
In the dilute instanton gas approximation, we assume that the average separation between two successive jumps is large compared to the width of a jump
\begin{equation}
    \Delta \tau \equiv \langle \tau_i - \tau_{i-1} \rangle \gg \omega^{-1}~.
\end{equation}
Then, the classical field configuration corresponding to $N$ successive jumps starting at $-v$ for $\tau\to-\beta/2$ can be approximately written as follows
\begin{equation}\label{eqn: approx multijump configuration}
    \phi_{N-\mathrm{jumps}} \approx  v (-1)^{N+1} \tanh \omega(\tau-\tau_1) \tanh \omega(\tau-\tau_2) \dots \tanh \omega(\tau-\tau_N)~,
\end{equation}
where $\tau_i$ is the location of the $i$-th jump.
Equation~\eqref{eqn: approx multijump configuration} is valid for both even and odd~$N$.
The above configurations are associated with their ``anti" type ones, which differ only by a minus sign. 
In Appendix~\ref{app: diga validity}, we discuss a condition for the dilute gas approximation to be valid.

The contribution to the partition function of the $N$-jumps configuration is given by
\begin{equation}\label{eqn: multi-instantons contribution}
    \left[K_1 \overline{K}_2 \ldots K_{N-1} \overline{K}_N \right] \approx \int_{-\frac{\beta}{2}}^{\frac{\beta}{2}} d\tau_1 \int_{\tau_1}^{\frac{\beta}{2}} d\tau_2 \dots \int_{\tau_{N-1}}^{\frac{\beta}{2}} d\tau_N [I]^N [v] = \frac{\beta^N}{N!} [I]^N [v]\,,
\end{equation}
where the integration region for the location of the jumps is fixed by an instanton always being followed by an anti-instanton and vice versa, 
a condition imposed by only having two degenerate minima.
Note that once again, we have isolated the contribution of fluctuations around the jumps from the contribution of fluctuations around the constant configuration.

The full partition function of the theory at quadratic order in the fluctuations is finally
\begin{equation}\label{eqn: partition function}
    Z_E \approx 2 \sum_{N=0}^\infty \frac{\beta^{2N}}{(2N)!} [I]^{2N} [v] = 2 [v] \cosh{\beta[I]} \,,
\end{equation}
where the factor of 2 is introduced by the two options $\phi(-\beta/2) = \pm v$.
Note that in Eq.~\eqref{eqn: partition function}, we only summed over configurations with an even number of jumps, which are the only configurations satisfying the periodic boundary conditions from Eq.~\eqref{eqn: E0 from Z}.
The ground state energy can then be expressed in terms of $[v]$ and $[I]$ using Eq.~\eqref{eqn: E0 from Z}, we find\footnote{Here we use that $[I]>0$, which can be readily seen from Eq.~\eqref{eqn:jump contribution}.}
\begin{equation}
    E_0 = - \lim_{\beta\to\infty} \frac{1}{\beta} \log [v] - [I] = E_{\mathrm{stat}} + E_{\mathrm{inst}}\,.
\end{equation}
As explained in Section~\ref{sec: casimir energy}, the contribution $E_\text{stat}$ of the static saddle points to the ground state energy can be identified with the Casimir energy once we renormalise it. 
The ratio of functional determinants appearing in $[I]$ is calculated in Section~\ref{sec: tunnelling}.

\section{Casimir energy}\label{sec: casimir energy}

The energy contribution of the static saddle points $E_{\mathrm{stat}}$ can be easily written down
\begin{align}
\label{eq:Estat0}
    E_{\mathrm{stat}} = & - \lim_{\beta\to\infty} \frac{1}{\beta} \log [v] \notag \\
    = & L^3 U_0 + \frac{1}{2} \lim_{\beta\to\infty} \frac{1}{\beta} \mathrm{tr} \log (-\partial_\tau^2 - \nabla^2 + m^2) \notag \\
    = & L^3 U_0 + \frac{1}{2}\sum_{\vec n \in \mathbb{Z}^3} \sqrt{\veck_n^2 + m^2}\,,
\end{align}
where $\veck_n = \frac{2\pi}{L} |\vec n|$. To arrive at the last equality, we have used
\begin{align}
    {\rm tr}\log (-\partial_\tau^2-\nabla^2+m^2)=\beta \int\frac{d \omega}{2\pi}\sum_{\vec{n}\in \mathbb{Z}^3} \log(\omega^2+\veck_n^2+m^2)
\end{align}
and performed the integral over $\omega$.
The energy $E_{\mathrm{stat}}$ is nothing but the vacuum energy~$E_0^{(3)}$ of a massive scalar field on a three-dimensional torus.
We can rewrite Eq.~\eqref{eq:Estat0} as
\begin{equation}
\label{eq:Estat2}
    E_{\rm stat}=L^3 U_0 + E_{\rm Casimir}^{(3)} + \frac{L^3}{2} \int_{\mathbb{R}^3} \frac{d^3\veck}{(2\pi)^3} \sqrt{\veck^2+m^2}\,,
\end{equation}
where we have adopted the standard definition of the Casimir energy~\cite{Bordag}. $E_{\rm Casimir}^{(3)}$ is finite.
The last divergent term will be eventually dealt with in renormalisation, see Eq.~\eqref{eq:Ueff} and below.
To compute the Casimir energy, we use the Abel-Plana formula
\begin{equation}
\label{eqn: Abel-Plana formula}
    \sum_{n=0}^\infty f(n) - \int_0^\infty d t\, f(t) = \frac{1}{2} f(0) + i \int_0^\infty d t\, \frac{f(i t) - f(-i t)}{e^{2\pi t} -1}
\end{equation}
which holds for any function $f(z)$ holomorphic in the region ${\rm Re}\, z \geq 0$ that vanishes fast enough towards infinity such that both the series and the integral are convergent. 
Thus, to use the formula, we must multiply the term in the sum $\sqrt{\veck_n^2 + m^2}$ by a regulator~$\eta (n)$, namely a smooth function that decays fast towards infinity and such that~$\eta(0)=1$. As explained in~\cite{Bordag}, we never need to state our regulator explicitly, as we will find a result that is completely independent of it.
The difference between the discrete and the continuum vacuum energy, yields the Casimir energy due to the compactness of the space
\begin{align}
    E_{\mathrm{Casimir}}^{(3)} (m; L,L,L) = \: & E_0^{(3)}(m;L,L,L) - E_\infty^{(3)}(m;L,L,L) \notag\\[2ex]
    = \: & \frac{1}{2}\sum_{\vec n \in \mathbb{Z}^3} \sqrt{\veck_n^2 + m^2} - \frac{L^3}{2} \int_{\mathbb{R}^3} \frac{d^3\veck}{(2\pi)^3} \sqrt{\veck^2+m^2} \,,
\end{align}
where our notation specifies that we are considering a three-torus of equal sizes of length~$L$.
The details of the calculations are found in Appendix~\ref{app: casimir energy technical details}. In the following, we outline the main steps.

It is easier to start by computing the Casimir energy on a one-dimensional space with periodic boundary conditions, namely a circle. The result for it is known (see ~\cite{Bordag})
\begin{equation}
    E_{\mathrm{Casimir}}^{(1)} (m; L) = \: - \frac{1}{\pi L} F_{(1,0)}(m L)\,,
\end{equation}
where the function $F_{(1,0)}(x)$ is defined as
\begin{align}
\label{eq:def-F10}
    F_{(1,0)}(x) = \int_x^\infty dy \, \frac{\sqrt{y^2 - x^2}}{e^{y}-1}~.
\end{align}
We highlight that $F_{(1,0)}(x)$ is a smooth function over the whole positive real line, and has the following asymptotic behaviours
\begin{align}
    F_{(1,0)} (0) = \: & \frac{\pi^2}{6} \,, \\
    F_{(1,0)} (mL\gg1) = \: & \sqrt{\frac{\pi}{2}} \sqrt{mL} e^{-mL} \left( 1 + \mathcal{O}((mL)^{-1}) \right)\,,
\end{align}
namely for very small lengths~$L$ the massless behaviour dominates the Casimir energy, while as~$L$ grows an exponential suppression kicks in.

Now we can use the recursive formula proven in Appendix~\ref{app: casimir energy technical details}, valid for the Casimir energy of a real massive scalar field on a $D$-dimensional torus of sizes~$a_1,\ldots ,a_D$ where~${D>1}$
\begin{align}\label{eqn: Casimir energy for generic D}
    E_{\mathrm{Casimir}}^{(D)} (m; a_1, \ldots,a_D ) = \: & - \frac{1}{\pi^D a_D} \left(\prod_{i=1}^{D-1} \frac{a_i}{a_D}\right) F_{(D,0)}(ma_D) \notag \\
    & + \sum_{n=-\infty}^\infty E_{\mathrm{Casimir}}^{(D-1)} \left( \sqrt{m^2 + \frac{4\pi^2n^2}{a_D^2}}; a_1, \ldots,a_{D-1} \right) \,,
\end{align}
where
\begin{align}
\label{eq:def-FD0}
F_{(D,0)} (x) = \: \left( \prod_{i=1}^{D-1} \int_0^\infty d y_i \right)  F_{(1,0)} \left( \scriptstyle{\sqrt{x^2 + \sum_{i=1}^{D-1}y_i^2} }\right) = \: \int_0^\infty d y F_{(D-1,0)} \left( \sqrt{x^2 + y^2} \right)\,.
\end{align}
Using the formula~\eqref{eqn: Casimir energy for generic D}, we find the Casimir energy for a real massive scalar field on a generic three-torus in Eq.~\eqref{eqn: casimir energy on a generic 3-torus}. Setting equal sizes $a_1=a_2=a_3=L$ we finally find\footnote{From now on, we write $E_{\mathrm{Casimir}}^{(3)}(m;L)$ instead of $E_{\mathrm{Casimir}}^{(3)} (m; L,L,L)$ since we will only be considering a torus with sides of equal length.}
\begin{equation}\label{eqn: casimir energy}
    E_{\mathrm{Casimir}}^{(3)} (m;L) = \: - \frac{1}{\pi^3 L} \left[ F_{(3,0)}(mL) + \pi F_{(2,1)}(1,mL) + \pi^2 F_{(1,2)} (1,1,mL) \right] \,,
\end{equation}
where the $F_{(i,j)}$ are defined as
\begin{subequations}
\begin{align}
    &F_{(i,1)} (r,x) = \: \sum_{n=-\infty}^\infty F_{(i,0)} (r \sqrt{x^2 + 4\pi^2n^2} )\,,\\
    &F_{(i,2)}(r_1,r_2,x) = \: \sum_{n=-\infty}^\infty F_{(i,1)} (r_1, r_2 \sqrt{x^2+4\pi^2n^2}) \,.
\end{align}
\end{subequations}

\section{The tunneling contribution}\label{sec: tunnelling}

We now turn to the calculation of the tunnelling (or instanton) contribution to the energy. In Section~\ref{sec: gas} we have derived
\begin{equation}\label{eqn: instanton energy}
    E_{\mathrm{inst}} = \: - [I] = \: - \sqrt{\frac{K}{2\pi}} e^{-K} \left(\frac{\sideset{}{'}\det \left(-\partial_\tau^2 - \nabla^2 + \frac{m^2}{2} \left(\frac{3\phi_K^2}{v^2} - 1\right)\right)}{\det (-\partial_\tau^2 - \nabla^2 + m^2)} \right)^{-\frac{1}{2}} \,.
\end{equation}
In this section, we compute the ratio of functional determinants. 
To do so, we employ the {\it resolvent method}, developed in~\cite{Baacke:1993jr,Baacke:1993aj,Baacke:1994ix,Baacke:2008zx}  and used in, e.g.,~\cite{Garbrecht:2015oea,Ai:2018guc,Ai:2020sru,Ai:2019fri}. The method states that the ratio of the determinants of two operators can be written as an integral of some modified Green's function. Explicitly
\begin{equation}\label{eqn: resolvent method}
    \log \frac{\det \hat A}{\det \hat B} = - \int_0^\infty ds \int d^4x \left( G_A(s;x,x) - G_B(s;x,x)\right)\,,
\end{equation}
where the modified Green's functions known as resolvents are defined through the differential equation
\begin{equation}
    \left(\hat A + s \right) G_A(s;x,y) = \delta^{(4)}(x-y)\,,
\end{equation}
and similarly for the operator $\hat B$.
To see Eq.~\eqref{eqn: resolvent method}, we write the modified Green's functions in their spectral representation, e.g.,
\begin{align}
    G_A(s;x,x')=\sum_n\frac{\phi^*_{A,n}(x) \phi_{A,n}(x')}{\lambda_{A,n}+s}\,, \qquad \text{where} \qquad \hat A \phi_{A,n}=\lambda_{A,n}  \phi_{A,n}
    \,.
\end{align}
Then 
\begin{align}
    \int_0^{S} d s\int d^4 x\, G_A(s;x,x)=-\sum_n\log \frac{\lambda_{A,n}}{\lambda_{A,n}+S}\,,
\end{align}
where we used that the eigenfunctions $\phi_{A,n}$ are normalised to one. Therefore,
\begin{align}
\label{eq: mode cancallation}
    - \int_0^{S} ds \int d^4x \left( G_A(s;x,x) - G_B(s;x,x)\right)=\sum_n\log\frac{\lambda_{A,n}}{\lambda_{B,n}}-\sum_n\log\frac{\lambda_{A,n}+S}{\lambda_{B,n}+S}\,.
\end{align}
When taking $S\rightarrow\infty$, the second term on the RHS vanishes.

\subsection{Finding the modified Green's function}

To compute the ratio of determinants appearing in Eq.~\eqref{eqn: instanton energy} via the resolvent method, we must find two modified Green's functions.
We will first evaluate the more complicated one, namely for the fluctuation operator around the instanton configuration. The equation reads
\begin{equation}
    \left(-\partial_\tau^2 - \nabla^2 + \frac{m^2}{2} \left(\frac{3\phi_K^2}{v^2} - 1\right) + s\right) G_K(s;x,x') = \delta^{(4)}(x-x')\,.
\end{equation}
Because of invariance under spatial translations, we can make the following ansatz
\begin{equation}
    G_K(s;x,x') = \frac{1}{L^3} \sum_{\vec n\in \mathbb{Z}^3} e^{-i\vec k_n \cdot (\vec x - \vec x')} F^{(K)}_n(s;\tau,\tau')\,,
\end{equation}
where the discretised momentum is $\vec k_n = \frac{2\pi}{L} \vec n$.
After changing to variable $u=\tanh \omega\tau$, the equation for the $\tau$-dependent part becomes
\begin{equation}
    \left( \frac{d}{d u} (1-u^2) \frac{d}{du} - \frac{\mu_n^2}{1-u^2} + 6 \right) \frac{m}{2} F^{(K)}_n(s;u,u') = - \delta(u-u')\,,
\end{equation}
where we have introduced the dimensionless quantity $\mu_n = 2 \sqrt{1 + \frac{\veck_n^2}{m^2} + \frac{s}{m^2}}$.

The solution to the equation above can be obtained in terms of associated Legendre functions $P_2^{\mu_n}(u)$ and $Q_2^{\mu_n}(u)$~\cite{Garbrecht:2018rqx,Ai:2019fri}.
It reads~\cite{Garbrecht:2015oea}
\begin{align} \label{eqn: non-subtracted kink green fnct}
    & F_n^{(K)} (s; u,u')  =  \frac{2}{m} \frac{1}{2\mu_n} \Big[ \theta(u-u') \left(\frac{1-u}{1+u}\right)^{\frac{\mu_n}{2}} \left(\frac{1+u'}{1-u'}\right)^{\frac{\mu_n}{2}} \notag \\ 
    &\qquad\quad \times \left(1-3\frac{(1-u)(1+\mu_n+u)}{(1+\mu_n)(2+\mu_n)}\right) \left( 1 - 3\frac{(1-u')(1-\mu_n + u')}{(1-\mu_n)(2-\mu_n)}\right) + ( u \leftrightarrow u') \Big]\,.
\end{align}
Note that the function above has a pole for $\mu_n=2$, namely for $n=0$ and $s=0$. This is not surprising: the instanton background breaks the time-translational symmetry of the theory, thus inducing a zero mode that appears as a simple pole in the physical Green's function~$G(0;x,x')$.
We need to define the subtracted Green's function on the space orthogonal to the zero mode.
Note that in the spectral decomposition, the modified Green's function is given by
\begin{equation}
    G(s;x,x') = \sum_{\lambda} \frac{\phi_\lambda(x)\phi_\lambda^*(x')}{\lambda+s} = \frac{\phi_0(x)\phi_0^*(x')}{s} + \sideset{}{'}\sum_\lambda \frac{\phi_\lambda(x)\phi_\lambda^*(x')}{\lambda+s}\,.
\end{equation}
Thus, to isolate the contribution of the zero mode~$\phi_0$ it is enough to look at the leading~$1/s$ behaviour in~$F_0^{(K)}$
\begin{equation}
    F_0^{(K)}(s;u,u') = \frac{3}{2m} \frac{(1-u^2)(1-u'^2)}{\mu_0^2-4} + \text{finite terms for }s\to0\,.
\end{equation}
By subtracting this term from expression~\eqref{eqn: non-subtracted kink green fnct}, we find the time-dependent part of the modified subtracted Green's function
\begin{align}
   F_n^{(K)\, \perp} (s; u,u') = & \frac{2}{m} \frac{1}{2\mu_n} \left[ \theta(u-u') \left(\frac{1-u}{1+u}\right)^{\frac{\mu_n}{2}} \left(\frac{1+u'}{1-u'}\right)^{\frac{\mu_n}{2}}\right. \notag \\[1.5ex] 
    & \times \left.\left(1-3\frac{(1-u)(1+\mu_n+u)}{(1+\mu_n)(2+\mu_n)}\right) \left(1 - 3\frac{(1-u')(1-\mu_n + u')}{(1-\mu_n)(2-\mu_n)}\right) + ( u \leftrightarrow u') \right] \notag \\[1.5ex]
    & - \delta_{n,0} \frac{3}{2m} \frac{(1-u^2)(1-u'^2)}{\mu_n^2-4}\,.
\end{align}
The coincident limit of the modified Green's function, needed for the resolvent method, is
\begin{align}
    G_K^\perp(s;x,x) = \: & \frac{2}{m L^3} \Bigg\{ \frac{1}{2\mu_0} + \frac{3(1-u^2)}{2\mu_0} \left( \frac{u^2}{\mu_0^2-1} - \frac{1}{2} \frac{1-u^2}{\mu_0+2} \right) + \notag \\
    & \qquad + \sideset{}{'}\sum_{\vec n \in \mathbb{Z}^3} \frac{1}{2\mu_n}\left( 1+  3(1-u^2) \sum_{\kappa=1}^2 (-1)^\kappa \frac{\kappa - 1 - u^2}{\mu_n^2 - \kappa^2} \right) \Bigg\} \,,
\end{align}
where the primed sum runs over $\mathbb{Z}^3\backslash \{0,0,0\}$.

We must now compute the modified Green's function for the fluctuation operator around the static saddle points. The differential equation reads
\begin{equation}
    (-\partial_\tau^2 - \nabla^2 + m^2 + s) G_v(s;x,x') = \delta^{(4)}(x-x')\,.
\end{equation}
Using the same ansatz and the same change of variable we performed above, we get to the equation for the time-dependent part
\begin{equation}
    \left( \frac{d}{d u} (1-u^2) \frac{d}{du} - \frac{\mu_n^2}{1-u^2}  \right) \frac{m}{2} F^{(v)}_n(s;u,u') = - \delta(u-u')\,,
\end{equation}
which can be solved in terms of associated Legendre functions of the first and second kind~$P_0^{\mu_n}(u)$ and~$Q_0^{\mu_n}(u)$. See for example \cite{Garbrecht:2018rqx}.
We have~\cite{Ai:2019fri}
\begin{align}
    F^{(v)}_n(s;u,u') = &  \frac{2}{m} \frac{1}{2\mu_n} \left[ \theta(u-u') \left(\frac{1-u}{1+u}\right)^{\frac{\mu_n}{2}} \left(\frac{1+u'}{1-u'}\right)^{\frac{\mu_n}{2}} + ( u \leftrightarrow u') \right]\,,
\end{align}
and in the coincident limit
\begin{equation}
    G_v(s;x,x) = \: \frac{2}{m L^3} \sum_{\vec n \in \mathbb{Z}^3} \frac{1}{2\mu_n}\,.
\end{equation}

\subsection{Computing the renormalised ratio of functional determinants}

We have all the ingredients we need to apply the resolvent method in Eq.\eqref{eqn: resolvent method}. We have
\begin{align}
    \log \frac{\sideset{}{'} \det S_E''[\phi_K]}{\det S_E''[v]} = & - \int_0^\infty ds \int d^3\vecx \int_{-\infty}^{\infty} d\tau \left( G_K^\perp (s;x,x) - G_v(s;x,x) \right) \notag\\[2ex]
    = & -\int_0^\infty ds \: \int_{-1}^1 \frac{2 du}{m(1-u^2)} \frac{2}{m} \Bigg\{ \frac{3(1-u^2)}{2\mu_0} \left( \frac{u^2}{\mu_0^2-1} - \frac{1}{2} \frac{1-u^2}{\mu_0+2} \right) + \notag\\
    & \qquad + \sideset{}{'}\sum_{\vec n \in \mathbb{Z}^3} \frac{1}{2\mu_n} 3(1-u^2) \sum_{\kappa=1}^2 (-1)^\kappa \frac{\kappa - 1 - u^2}{\mu_n^2 - \kappa^2} \Bigg\} \notag\\[2ex]
    = & - \log 3 + \frac{4}{m^2} \int_0^\infty ds \: \frac{1}{\mu_0 (\mu_0 + 2)} \notag\\
    & - 2\sideset{}{'}\sum_{\vec n \in \mathbb{Z}^3} \log \frac{3m^2 + 2\veck_n^2 + 3 m \sqrt{\veck_n^2 + m^2}}{|\veck_n| \sqrt{4\veck_n^2+3m^2}} \,.
\end{align}
The expression above still contains a divergent integral
\begin{equation*}
     \frac{4}{m^2}\int_0^S ds \: \frac{1}{\mu_0 (\mu_0 + 2)} = - \log 4m^2 + \log S + \mathcal{O}(S^{-1}) \to \infty \,.
\end{equation*}
This arises because of the mismatch in the number of eigenmodes between the two operators in the ratio, a consequence of having subtracted the zero mode from the fluctuation operator around the kink. Specifically, this can be seen from Eq.~\eqref{eq: mode cancallation}. Removing the zero mode from $G_A$ would lead to a divergent contribution $\log (\lambda_{B,0}+S)\sim \log S$ for $S\rightarrow\infty$ from the second term on the RHS.
Therefore, this divergence is a mere artefact of removing a zero mode in the resolvent-formula for the functional determinants Eq.~\eqref{eqn: resolvent method} and should be dropped out.
We find an expression for the un-renormalised logarithm of the ratio of determinants
\begin{equation}
\begin{aligned}
\label{eq:log-det-kink}
    \frac{1}{2} \log \frac{\sideset{}{'} \det S_E''[\phi_K]}{\det S_E''[v]} = & - \frac{1}{2}\log 12m^2 - \sideset{}{'}\sum_{\vec n \in \mathbb{Z}^3} \log \frac{3m^2 + 2\veck_n^2 + 3 m \sqrt{\veck_n^2 + m^2}}{|\veck_n| \sqrt{4\veck_n^2+3m^2}} \,.
\end{aligned}
\end{equation}
We can identify this quantity as the one-loop correction to the 1PI effective action~\cite{Jackiw:1974cv} around the kink configuration, having subtracted the correction around the static one.

The divergences from the last term in Eq.~\eqref{eq:Estat2} and in~\eqref{eq:log-det-kink} can only be removed via renormalisation of the one-loop effective action $\Gamma^{(1)}[\phi]$.
With the counter-terms, we have explicitly
\begin{align}\label{eqn: difference effective actions}
    \Gamma^{(1)}[\phi_K] & - \Gamma^{(1)}[v] = \: S_E[\phi_K] - S_E[v] + \frac{1}{2}\log \frac{\det S''_E[\phi_K]}{\det S''_E[v]} + S_{c.t.}[\phi_K]-S_{c.t.}[v] \notag \\
    = \: & K + \frac{1}{2}\log \frac{\det S''_E[\phi]}{\det S''_E[v]} + \int_0^\infty d\tau \int d^3 \vecx \left( \frac{\delta m^2}{2} (\phi_K^2 - v^2) + \frac{\delta\lambda}{4!} (\phi_K^4 - v^4) \right) \,.
\end{align}
Note that once the counter-terms have been added, all the coupling constants, including the mass and $U_0$, are the renormalised ones and are to be intended as such in what follows. For clarity, we still use the same notations as for the bare quantities.
Then the instanton contribution to the energy is 
\begin{equation}
    E_{\mathrm{inst}} = \: -e^{-\left(\Gamma^{(1)}[\phi_K] - \Gamma^{(1)}[v] \right)}\,.
\end{equation}
Since the zero-point energy $U_0$ is field independent, it is cancelled out in the difference. The UV divergences are sensitive to neither the background nor the topology of space. Therefore, we can find $\delta m^2$ and $\delta \lambda$ by simply renormalising the one-loop effective potential in the infinite volume limit that is obtained from the effective action for constant field backgrounds,
\begin{subequations}
\begin{align}
    &\qquad\qquad\qquad\left.\Gamma^{(1)}[\phi]\right|_{\phi \text{ is constant}}=\int d^4 x\, U_{\rm eff}(\phi)\,,\\[1.5ex]
    & \Rightarrow U_{\rm eff}(\phi)=U(\phi)+\frac{1}{2}\int\frac{d^3\veck}{(2\pi)^3}\sqrt{\veck^2+U''(\phi)} +\frac{\delta m^2}{2}\phi^2 +\frac{\delta\lambda}{4!} \phi^4+\delta U_0\,. \label{eq:Ueff}
\end{align}
\end{subequations}
The integral above is divergent, and we need to choose a regularisation procedure that we will use to regulate both the integral in the continuum limit and its corresponding series. We use a smooth-regulator
\begin{equation}
    \eta \left(\frac{|\veck|}{\Lambda} \right) = \: e^{-\frac{|\veck|^2}{\Lambda^2}}\,,
\end{equation}
where $\Lambda$ is some large cut-off scale.
With this, the integral reads
\begin{align}
     &\frac{1}{2} \int\frac{d^3\veck}{(2\pi)^3} \sqrt{\veck^2+U''(\phi)} \, e^{-\frac{\veck^2}{\Lambda^2}}\notag\\
     &\qquad=  \frac{[U''(\phi)]^2}{128\pi^2} \left[ \frac{16\Lambda^4}{(U''[\phi])^2} + \frac{8\Lambda^2}{U''(\phi)} + 1 + 2\gamma_E + 2\log \frac{U''(\phi)}{4\Lambda^2} \right] + \mathcal{O}\left(\frac{\sqrt{U''(\phi)}}{\Lambda}\right) \,.
\end{align}
We choose the following renormalisation conditions
\begin{subequations}
\begin{align}
    U_{\mathrm{eff}}(\phi)\rvert_{\phi=\pm v} = \: & U_0\,, \label{eq:renormalisation-cond1}\\
    \frac{d^2 U_{\mathrm{eff}}(\phi)}{d\phi^2}\rvert_{\phi=\pm v} = \: & m^2 \,,\\
    \frac{d^4 U_{\mathrm{eff}}(\phi)}{d\phi^4}\rvert_{\phi=\pm v} = \: &\lambda \,.
\end{align}
\end{subequations}
Imposing the renormalisation conditions fixes the counter-terms
\begin{subequations}
\begin{align}
    \delta U_0 = & -\frac{m^4}{256\pi^2}\left(143 + \gamma_E - \frac{8\Lambda^2}{m^2} + \frac{32\Lambda^4}{m^4} - \log\frac{4\Lambda^2}{m^2} \right)\,,\label{eqn: V0 counterterm}\\
    \delta m^2 = & - \frac{\lambda m^2}{64\pi^2} \left(-31 -\gamma_E + \frac{4\Lambda^2}{m^2} + \log\frac{4\Lambda^2}{m^2}\right)\,, \label{eqn: mass counterterm}\\
    \delta \lambda = & - \frac{3\lambda^2}{32\pi^2} \left(5 + \gamma_E - \log \frac{4\Lambda^2}{m^2}\right) \label{eqn: coupling counterterm}\,.
\end{align}
\end{subequations}
With the above renormalisation procedure, the divergent term in Eq.~\eqref{eq:Estat2} has essentially been regularised by the counter-term $\delta U_0$, leaving the renormalised, physical $U_0$ through the condition~\eqref{eq:renormalisation-cond1}.

The contribution of the counter-terms to the effective action reads
\begin{align}
    S_{c.t.}[\phi_K] - S_{c.t.}[v] = \: & \int_{-\infty}^\infty d\tau \int d^3\vecx \left( \frac{\delta m^2}{2} (\phi_K^2-v^2) + \frac{\delta\lambda}{4!} (\phi_K^4-v^4) \right) \notag\\
    = \: & - 2 m^3L^3 \left(\frac{3\delta m^2}{m^2\lambda} + \frac{\delta\lambda}{\lambda^2} \right) \notag\\
    = \: & + \frac{3 m^3L^3}{32\pi^2} \left( -21 + \gamma_E + \frac{4\Lambda^2}{m^2} - \log\frac{4\Lambda^2}{m^2}\right)\,.
\end{align}
The resulting expression for the instanton contribution to the ground state energy is
\begin{align}\label{eqn: kink energy}
    E_{\mathrm{inst}} = & - \sqrt{12m^2} \sqrt{\frac{m^3 L^3}{\pi\lambda}} \exp \bigg\{ - \frac{2m^3L^3}{\lambda} \notag \\
    & \qquad + \sideset{}{'}\sum_{\vec n \in \mathbb{Z}^3} e^{-\frac{k_n^2}{\Lambda^2}} \log \frac{3m^2 + 2k_n^2 + 3 m \sqrt{k_n^2 + m^2}}{k_n \sqrt{4k_n^2+3m^2}} \notag \\
    & \qquad - \frac{3 m^3L^3}{32\pi^2} \left( -21 + \gamma_E + \frac{4\Lambda^2}{m^2} - \log\frac{4\Lambda^2}{m^2}\right) \bigg\} \,.
\end{align}
We can check that these counter-terms correctly renormalise the series by considering the continuum limit of the latter, and extracting its UV divergences.
We find
\begin{align}
    \sideset{}{'}\sum_{\vec n \in \mathbb{Z}^3} e^{-\frac{\veck_n^2}{\Lambda^2}} & \log \frac{3m^2 + 2\veck_n^2 + 3 m \sqrt{\veck_n^2 + m^2}}{|\veck_n| \sqrt{4\veck_n^2+3m^2}} \notag\\[2ex]
    \longrightarrow \: & L^3 \int \frac{d^3\veck}{(2\pi)^3} e^{-\frac{\veck^2}{\Lambda^2}} \log \frac{3m^2 + 2\veck^2 + 3 m \sqrt{\veck^2 + m^2}}{|\veck| \sqrt{4\veck^2+3m^2}}  \notag\\[2ex]
    = \: & \frac{L^3}{2\pi^2} \int_\epsilon^\infty d|\veck|\: \veck^2  e^{-\frac{\veck^2}{\Lambda^2}} \left( \frac{3m}{2|\veck|} - \frac{3m^3}{8|\veck|^3} + \mathcal{O}\left(\frac{m^5}{|\veck|^5}\right) \right) \notag\\[2ex]
    = \: & \frac{3m^3L^3}{32\pi^2} \left( \frac{4\Lambda^2}{m^2} - \log \frac{4\Lambda^2}{m^2} \right) + \text{finite terms} \,,
\end{align}
which are indeed cancelled by the counter-term contribution in Eq.~\eqref{eqn: kink energy}.

\begin{figure}
    \centering
    \begin{subfigure}[b]{0.45\textwidth}
        \centering
        \includegraphics[width=\textwidth]{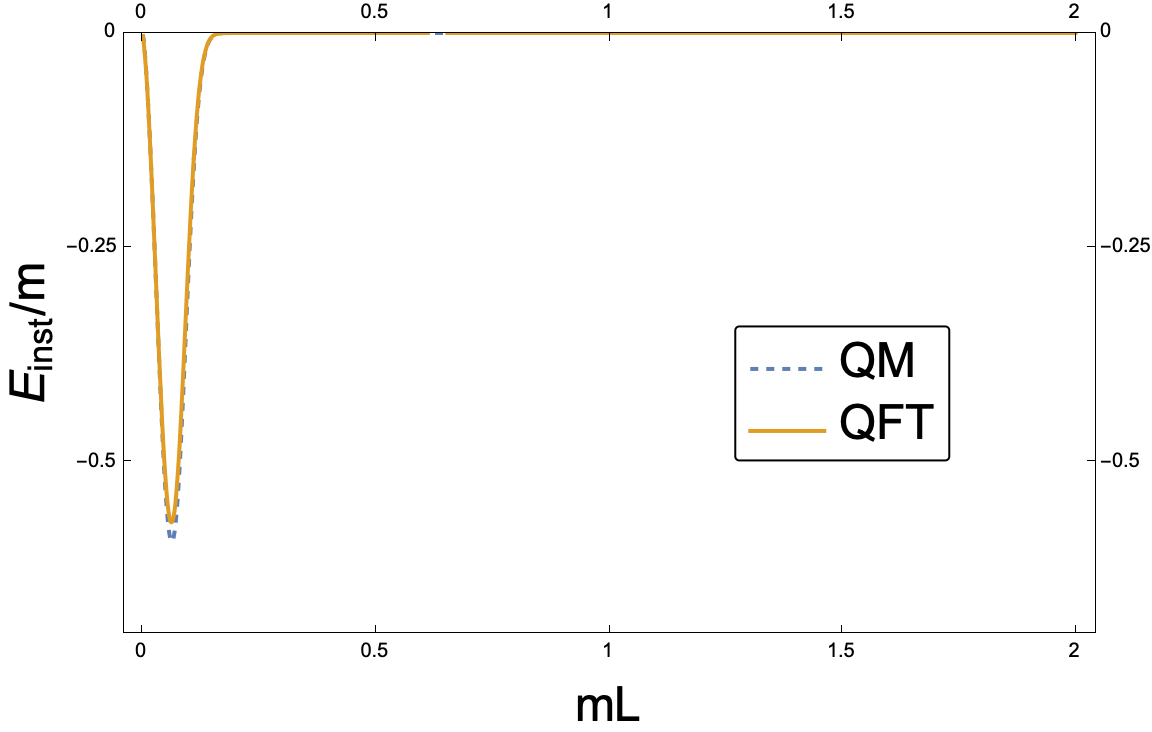}
        \caption{$\lambda=0.001$}
    \end{subfigure}
    \hfill
    \begin{subfigure}[b]{0.45\textwidth}
        \centering
        \includegraphics[width=\textwidth]{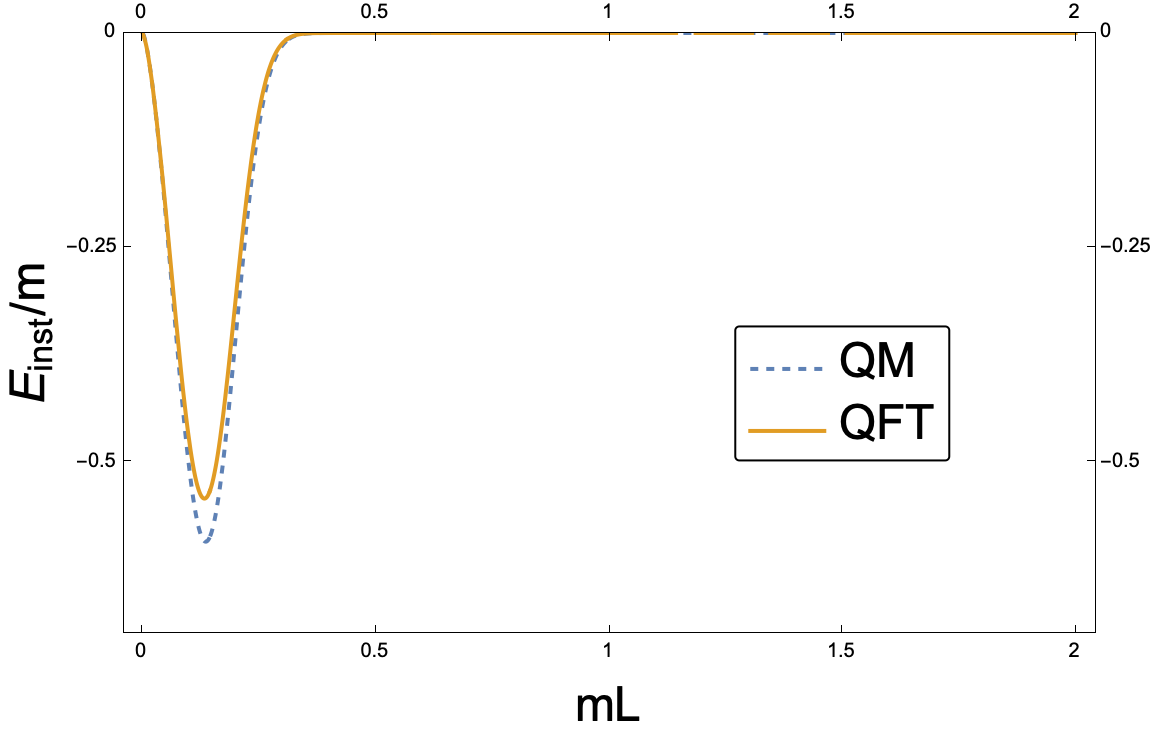}
        \caption{$\lambda=0.01$}
    \end{subfigure}
    \\[2ex]
    \begin{subfigure}[b]{0.45\textwidth}
        \centering
        \includegraphics[width=\textwidth]{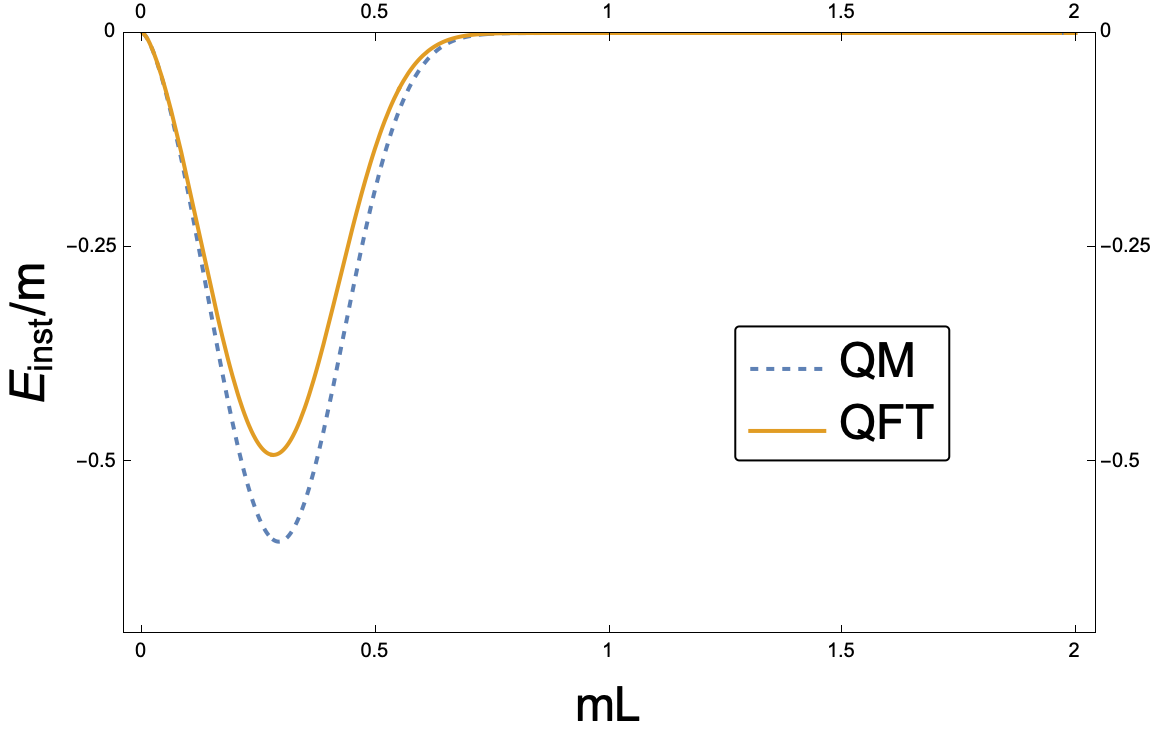}
        \caption{$\lambda=0.1$}
    \end{subfigure}
    \hfill
    \begin{subfigure}[b]{0.45\textwidth}
        \centering
        \includegraphics[width=\textwidth]{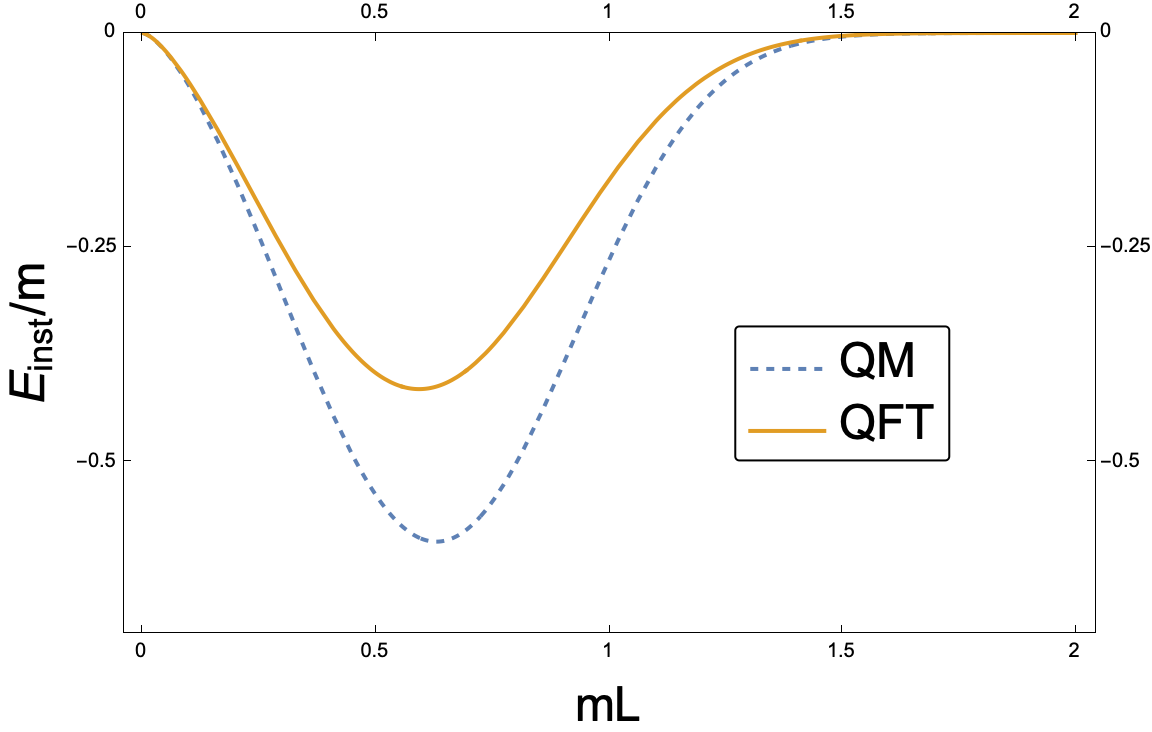}
        \caption{$\lambda=1$}
    \end{subfigure}
    \caption{Comparison between the quantum mechanical approximation and the full QFT result of $E_{\rm inst}$, as defined in Eq.~\eqref{eqn: QM approximation to the kink energy} and Eq.~\eqref{eqn: kink energy} respectively.}
    \label{fig: qm vs qft}
\end{figure}

The result in Eq.~\eqref{eqn: kink energy} includes the full one-loop contribution of the fluctuations around the jump. In Refs.~\cite{Alexandre:2023iig,Alexandre:2023pkk}, the instanton contribution to the ground state energy is calculated by only keeping homogeneous fluctuations. This can be obtained from the result in Eq.~\eqref{eqn: kink energy} by neglecting the contributions from $\vec n \neq \vec 0$ fluctuations, namely
\be\label{eqn: QM approximation to the kink energy}
    E^{(\mathrm{hom})}_{\mathrm{inst}}=-\sqrt{12 m^2}\sqrt{\frac{m^3L^3}{\pi \lambda}}\exp\left\{ -\frac{2 m^3 L^3}{\lambda}\right\}\,.
\ee
The calculation of this result resembles one of a tunnelling problem in Quantum Mechanics (QM).
Therefore, we will refer to Eqs.~\eqref{eqn: QM approximation to the kink energy} and~\eqref{eqn: kink energy} as QM and QFT results, respectively.
The inhomogeneous fluctuations yield additional one-loop corrections, which we expect to be perturbatively small compared to the QM result, at least for small coupling~$\lambda$. 
In Figure~\ref{fig: qm vs qft}, we plot the QM and the QFT results.
As expected, the smaller $\lambda$ the smaller the change in the instanton energy. In particular, for $\lambda \ll 1$, the QM result approximates the full QFT result really well. This justifies the approximations made in Refs.~\cite{Alexandre:2023iig,Alexandre:2023pkk}, where the spatial dependence of the quantum fluctuations was ignored.
Interestingly, for larger $\lambda$ the peak of the energy in the QFT result shrinks and moves to smaller values of $mL$ when compared to the QM one.

\subsection{Comparing with the Casimir energy}

\begin{figure}
    \centering
    \begin{subfigure}[b]{0.7\textwidth}
        \centering
        \includegraphics[width=\textwidth]{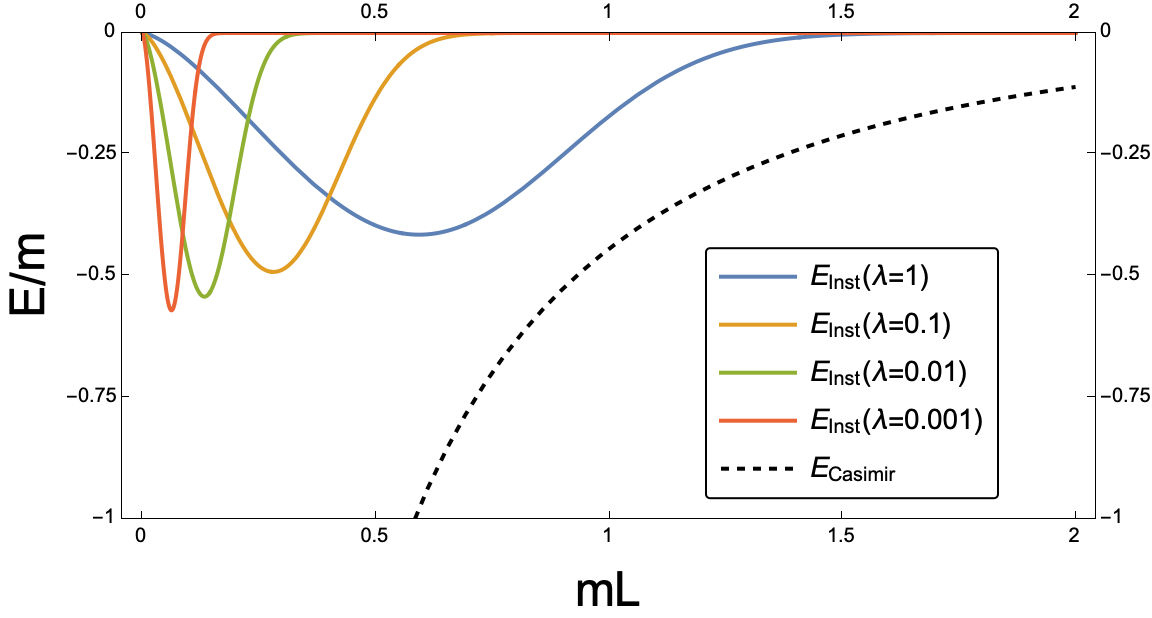}
        \caption{}
    \end{subfigure}
    \\[2ex]
    \begin{subfigure}[b]{0.75\textwidth}
    \centering
    \includegraphics[width=\textwidth]{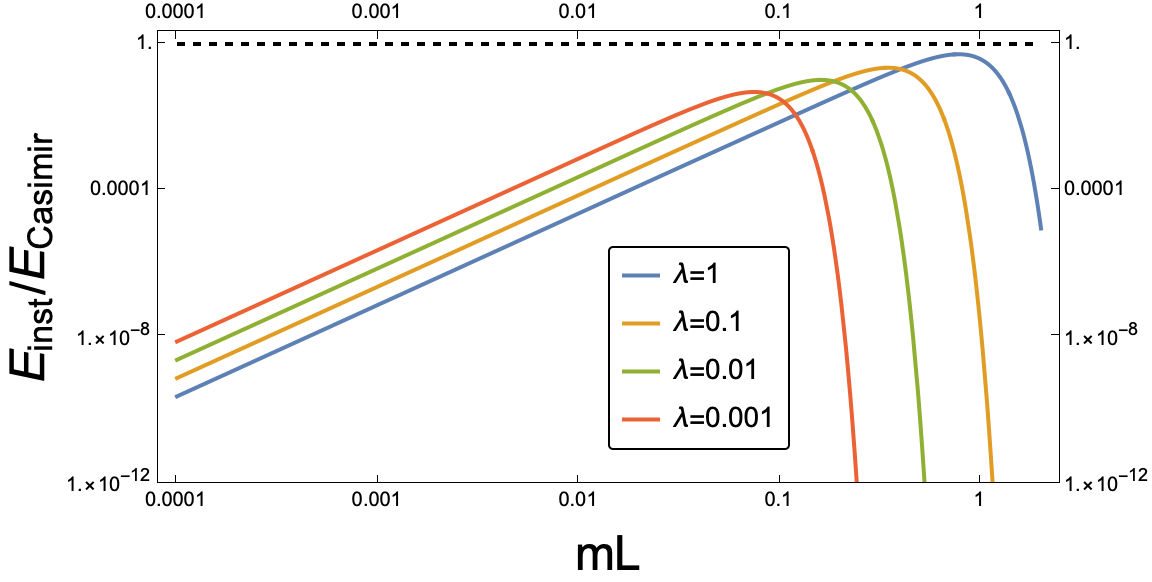}
    \caption{}
    \end{subfigure}
    \caption{Comparison of the contributions to the ground state energy as defined in Eq.~\eqref{eqn: casimir energy} and~\eqref{eqn: kink energy} for various values of~$\lambda$.}
    \label{fig: casimir vs tunnelling energy}
\end{figure}

We can now compare the relative contribution of the Casimir effect and the tunnelling to the ground state energy. We show the comparison in Figure~\ref{fig: casimir vs tunnelling energy}.
At first glance, we see that the Casimir energy is generally larger than the instanton one. However, for $mL\sim 1$ the latter can become of the same order of magnitude as the former, as can be seen in the bottom panel of Figure~\ref{fig: casimir vs tunnelling energy}.

\begin{figure}
    \centering
    \begin{subfigure}[b]{0.76\textwidth}
        \centering
        \includegraphics[width=\textwidth]{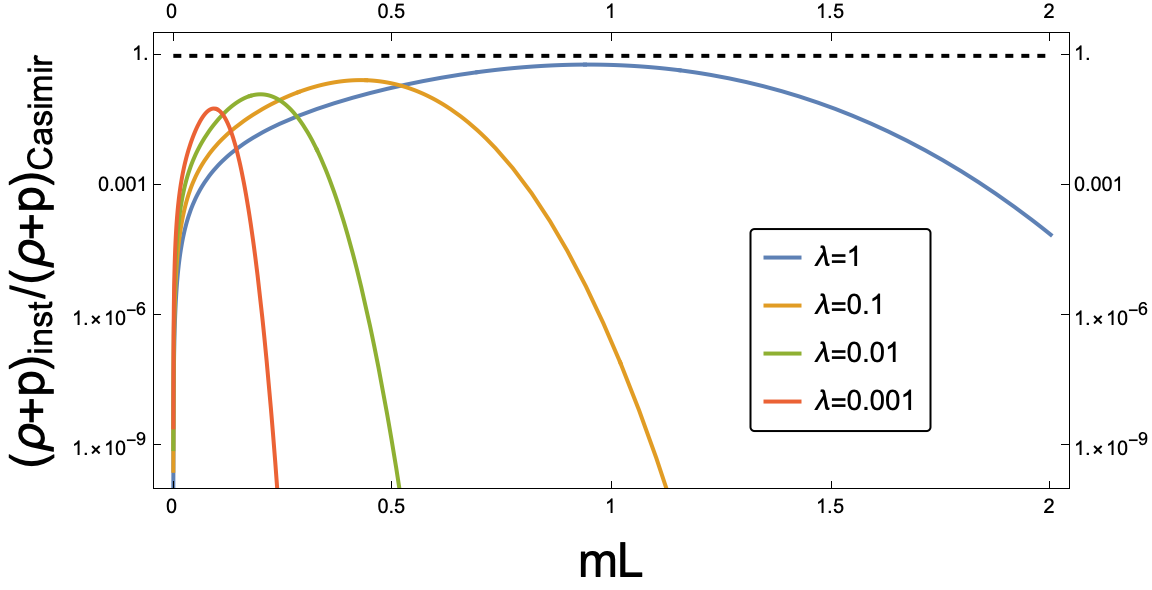}
        \caption{}
    \end{subfigure}
    \\[2ex]
    \begin{subfigure}[b]{0.7\textwidth}
        \centering
        \includegraphics[width=\textwidth]{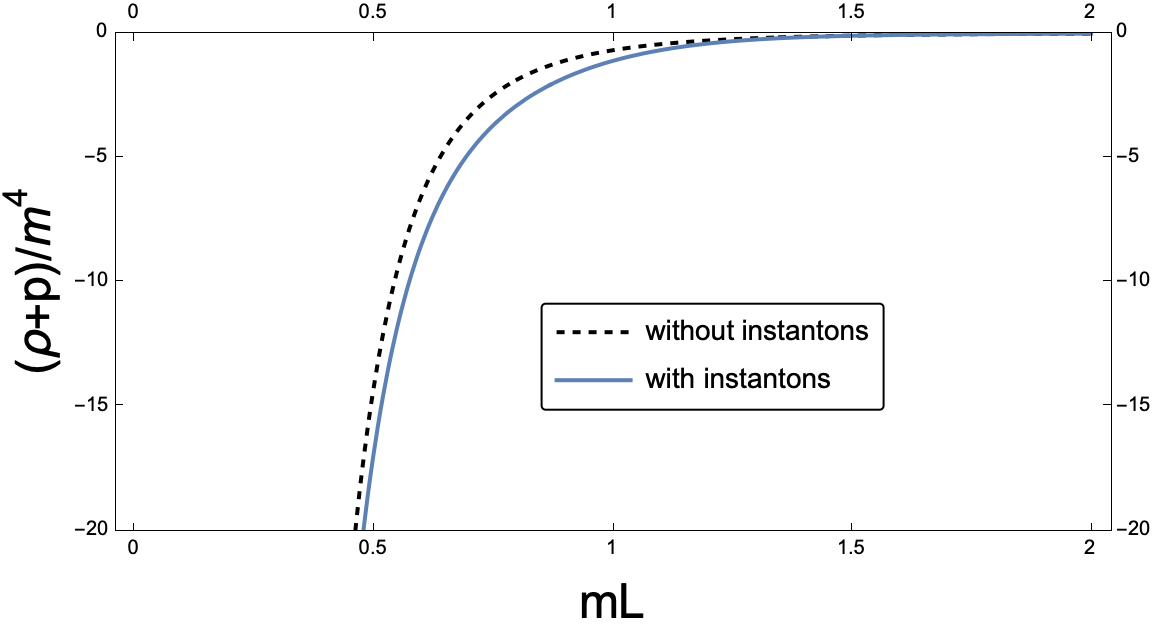}
        \caption{}
    \end{subfigure}
    \caption{Comparison of the contributions to $\rho+p$ as defined in Eq. \eqref{eqn: energy density plus pressure} from the tunnelling and Casimir effects, respectively. Panel~(a): the ratio of the contribution of the instantons over the contribution of the Casimir effect is shown for various values of $\lambda$. Panel~(b): the sum of energy density and pressure is shown with and without instanton effects for $\lambda=1$.}
    \label{fig: T00 casimir vs tunnelling}
\end{figure}

Because of its relevance for violating the NEC in Eq.~\eqref{eqn: NEC}, we are primarily interested in the sum of energy density and pressure, which in the thermodynamic limit can be written as
\begin{equation}\label{eqn: energy density plus pressure}
    \rho + p = \: \frac{E_0}{V} - \frac{\mathrm{d} E_0}{\mathrm{d} V}\,,
\end{equation}
where $E_0$ is the renormalised ground state energy and $V=L^3$ is the spatial volume. Specifically, the energy $E_0$ reads
\be
E_0=L^3U_0+E_{\textrm{Casimir}}^{(3)}+E_{\textrm{inst}}\,,
\ee
where $E_{\textrm{Casimir}}^{(3)}$ is given in Eq.~\eqref{eqn: casimir energy} and $E_{\textrm{inst}}$ is given in Eq.~\eqref{eqn: kink energy}.
In Figure~\ref{fig: T00 casimir vs tunnelling}, we plot the relative contribution of the Casimir effect and the instantons to $\rho+p$. The top panel clearly shows that for large enough coupling $\lambda$ the effect of instantons becomes of the same order as the Casimir effect. 
In the bottom panel $\rho+p$ is plotted, both ignoring and including instanton effects for $\lambda=1$. 
The instanton effects result in a steeper curve, thus making the NEC violation stronger.

\section{Relevance to Cosmology} 
\label{sec: applications}

Singularity theorems in general relativity show that singularities are inevitable under very general circumstances~\cite{Hawking:1970zqf}. However, they have in their assumptions some restrictions on the energy-momentum tensor for matter fields $T_{\mu \nu}$. In particular, it is assumed to satisfy the NEC $T_{\mu \nu}\ell^\mu \ell ^\nu \geq 0$ where $\ell^{\mu}$ is a null vector. For a perfect fluid, the NEC reduces to
\be\label{eqn: NEC}
\rho+p\geq 0\,,
\ee
where $\rho$ and $p$ are the energy density and the pressure of the fluid respectively. Therefore, identifying new NEC-violating mechanisms within quantum field theory holds particular significance in cosmology since the classical singularity theorems do not apply, and the cosmological singularity can be potentially avoided. 

The finite volume effects described in this work could naturally induce a semi-classical bounce when the effective theory is coupled to gravity in a spatially compact universe, such as the one considered here.

Recently, there have been discussions that Cosmic Microwave Background temperature and polarization data from Planck favour a spatially closed Universe~\cite{DiValentino:2019qzk,2021PhRvD.103d1301H}, although such preference disappears when the Planck data are combined with other measurements~\cite{Vagnozzi:2020rcz, Vagnozzi:2020dfn,Dhawan:2021mel}. For an FLRW universe, the observational data only constrains the curvature parameter~$\Omega_{K}$, defined as~${\Omega_{K}=-K/(a_0^2H_0^2)}$,\footnote{ Here the FLRW metric is: \( {\rm d}s^2=-{\rm d}t^2+a^2(t)\left(\frac{{\rm d} r^2}{1-Kr^2}+r^2{\rm d}\Omega_2\right)~. \)} and currently all cases $K=0,\pm 1$ remain possible.\footnote{ Ref.~\cite{Vagnozzi:2020dfn} finds $\Omega_K = -0.0054 \pm 0.0055$.} Although in the present work we assume the space to be a flat three-torus ($K=0$) for simplicity, the analysis can be generalised to a three-sphere ($K=1$) and is left for future work.

The analysis of the Casimir effect in a cosmological context was introduced for the first time in Ref.~\cite{Zeldovich:1984vk}. Since then, several authors studied its possible effects in cosmology (see Ref.~\cite{Bordag} for a review). For example, in Ref.~\cite{Herdeiro:2005zj} the backreaction problem is discussed when considering the Casimir energy of a massless field, and static solutions to the Friedmann equations are derived for a closed universe with positive curvature. Other works based on the Casimir effect, have considered the possibility of inducing an accelerating universe~\cite{Szydlowski:2007bg,Godlowski:2007gx,Herdeiro:2005zj}, but did not consider tunnelling. We also mention here works involving the scalar Casimir effect that are related to non-trivial universe topology.
In Refs.~\cite{Saharian:2007ts,Saharian:2010nep,Pavlov:2020utn}, the vacuum energy for a scalar field is evaluated in either a de Sitter universe or an FLRW universe, with one or more compact extra dimensions. Also, the stabilisation of the Einstein universe via the Casimir force is discussed in Ref.~\cite{Herdeiro:2007eb}. 

The effect of quantum tunnelling in cosmology was recently analysed in Ref.~\cite{Alexandre:2023pkk}. Ignoring the Casimir energy and the space-dependence of quantum fluctuations about the instanton, it is found that tunnelling can indeed dynamically generate an effective equation of state that supports a cosmological bounce when coupled to gravity. The energy density and the pressure of the effective ground state were obtained upon assuming that the tunnelling rate is large compared to the expansion rate of the universe,  
\be \label{eq:I0}
|H|\equiv \Big|\frac{\dot a}{a}\Big|\ll [I^{(0)}]\,,\quad \text{where}\quad [I^{(0)}]=  \sqrt{12m^2} \sqrt{\frac{m^3 L^3}{\pi\lambda}} \exp \bigg\{ - \frac{2m^3L^3}{\lambda}\bigg\}~,
\ee
which allows the calculation of quantum fluctuations for a constant scale factor. The time-dependence of the latter is then restored when coupling the effective theory to gravity.
In this context, the energy density and the pressure of the ground state fluid read
\begin{subequations}
    \begin{align}
       \rho^{(0)}(t)&=\frac{E_{0}^{(0)}(t)}{L(t)^3}= U_0 - \sqrt{12m^2}\sqrt{\frac{m^3}{\pi \lambda L(t)^3}}e^{-\frac{2 m^3}{\lambda} L(t)^3}\, ,\\
p^{(0)}(t)&=-\frac{1}{3 L(t)^2}\frac{\textrm{d}E_{0}^{(0)}(t)}{\textrm{d}L(t)}\notag\\
&=-U_0+\sqrt{12 m^2}\sqrt{\frac{m^3}{\pi \lambda L(t)^3}}\left(\frac{1}{2}-\frac{2m^3 L(t)^3}{\lambda}\right)e^{-\frac{2 m^3}{\lambda} L(t)^3}\, , 
    \end{align}
\end{subequations}
where $L(t)=a(t) L_0$, with $L_0$ being an arbitrary ``length-cell". It can be shown (see Ref.~\cite{Alexandre:2023pkk}) that the NEC is violated for any value of the comoving volume $a^3L_0^3$, and therefore a bouncing solution is obtained in FLRW independently of the initial conditions as long as~$H(t_0)<0$.  

An advantage of these finite-volume mechanisms is that they intrinsically favour expansion over contraction since they are exponentially suppressed after a period of expansion (i.e., for large volumes). 
Also, they do not rely on modified gravity or exotic matter and are dynamically generated by a shrinking co-volume.

Including the Casimir energy and one-loop corrections to the instanton jump should not modify the qualitative results in Ref.~\cite{Alexandre:2023pkk} since the NEC is still violated for any value of the comoving volume, which means that we expect to find a bouncing solution to the Friedmann equations if we start from a contracting phase. However, quantitative results can be substantially different since the Casimir contribution is very large. In Ref.~\cite{Alexandre:2023bih}, an estimation for the size of the universe at the bounce $L_{\textrm{bounce}}$ was given. Given that the NEC is always violated, the cosmic bounce will happen when 
\be
H^2=\frac{\dot a^2}{a^2}=0\, , \qquad \to \qquad
E_{0}=L^3U_0+E_{\textrm{Casimir}}^{(3)}+E_{\textrm{inst}}=0\,.
\ee
From this equation, we can estimate $L_{\textrm{bounce}}$. For example, in the simplified model analysed in Refs.~\cite{Alexandre:2023pkk, Alexandre:2023bih}, where  $E_{\textrm{cas}}^{(3)}=0$ and $E_{\rm inst}=E_{\rm inst}^{(\mathrm{hom})}$ with $E_{\rm inst}^{(\mathrm{hom})}$ given in Eq.~\eqref{eqn: QM approximation to the kink energy}, it was found $L_{\textrm{bounce}}\simeq 10^{-5}$ metres for  $U_0=\kappa^{-1}\Lambda=10^{-122}\,\ell_p^{-4}$, $m=10^{-3}$ eV and $\lambda \sim 1$. 
Repeating the computation with the one-loop ground state energy~$E_0$ as obtained in this work, we obtain $L_{\textrm{bounce}}\simeq 10^{-2}$ metres instead. These conclusions might be modified with space in the form of a three-sphere, leading to different relations between the size of the universe at the bounce, the present cosmological constant and the mass of the scalar field experiencing tunnelling.

\section{Conclusion and outlook}\label{sec: conclusions}

A quantum field system in a finite volume and an infinite volume can have very different behaviours. It is well known that a finite volume can generally induce a negative contribution to the energy of the system, known as the Casimir energy. When considering a field potential with degenerate minima, the finite volume allows tunnelling between those degenerate minima, which would otherwise be prohibited in the infinite-volume limit. The mentioned tunnelling also induces a negative ground state energy when the potential of the degenerate minima is normalised to zero and meanwhile restores the symmetry that is classically broken by the degenerate minima. In this paper, we have extensively studied both the Casimir and tunnelling effects for a scalar field theory with a double-well potential in a three-torus.

We have computed the contributions to energy from the Casimir and the tunnelling effects at the one-loop level. We derived analytic expressions for the Casimir energy for an arbitrary $D$-dimensional torus using the Abel-Plana formula. The tunnelling between the degenerate minima is described by a homogeneous but time-dependent Euclidean bounce solution. At the one-loop level, one has to integrate the fluctuations about the bounce solution as well as the classical degenerate vacua, which then need to be properly renormalised. This is achieved through the resolvent method~\cite{Baacke:1993jr,Baacke:1993aj,Baacke:1994ix,Baacke:2008zx} and renormalisation via the one-loop effective potential. 

The finite volume constraint has remarkable energetic effects on the confined system, allowing the violation of the NEC.
In addition to the known Casimir effect, we show that the tunnelling effect also contributes to NEC violation, extending the results in Ref.~\cite{Alexandre:2023pkk}. Depending on the coupling constant, the tunnelling effect can be of a similar order of magnitude as the Casimir effect. The violation of the NEC could induce a cosmic bounce~\cite{Alexandre:2023bih}.

The present work may have several generalisations.
The presence of fermions (satisfying anti-periodic boundary conditions) could decrease or cancel the Casimir contribution from the scalar field, provided the appropriate number of degrees of freedom is considered.\footnote{The extension of the Casimir effect to Supergravity in relation to Cosmology has been studied in Ref.~\cite{Goncharov:1987tz}.}
In this situation, tunnelling would become the main origin of NEC violation.
Other geometries or topologies could be considered, for example a three-sphere instead of a three-torus, which may be relevant to Cosmology.

\acknowledgments

The work of W.Y.A. and J.A. is supported by the Engineering and Physical Sciences Research Council (grant No. EP/V002821/1).
The work of J.A. and S.P. is supported by the Leverhulme Trust (grant No. RPG-2021-299). 
J.A. is also supported by the Science and Technology Facilities Council (grant No.  STFC-ST/X000753/1). 
For the purpose of Open Access, the authors have applied a CC BY public copyright licence to any Author Accepted Manuscript version arising from this submission.

\newpage

\appendix

\section{Validity of the dilute instanton gas approximation}\label{app: diga validity}

\begin{figure}
    \centering
    \includegraphics[width=0.7\textwidth]{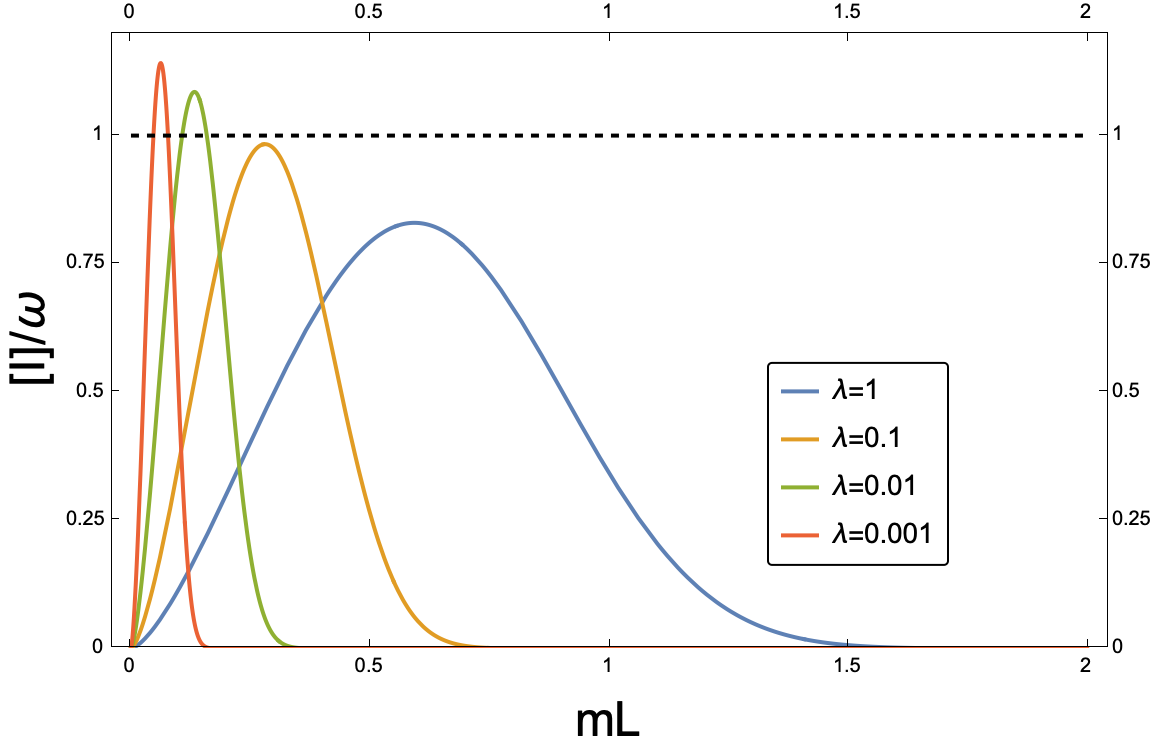}
    \caption{The instanton rate $[I]$ in units of $\omega$ for various values of the self-coupling $\lambda$.}
    \label{fig: diga validity}
\end{figure}

The dilute instanton gas approximation is valid if consecutive jumps are widely separated, namely if the average separation is much larger than the jump width
\begin{equation}
    \Delta\tau  \gg \omega^{-1}\,.
\end{equation}
The average separation is given by the total length of the Euclidean time interval divided by the average number of jumps in said interval
\begin{equation}
    \Delta \tau = \frac{\beta}{\langle N \rangle}\,.
\end{equation}
The average number of jumps is 
\begin{equation}
    \langle N \rangle = 
    \left( \sum_{N=0}^\infty \frac{([I]\beta)^N}{N!}\right)^{-1} \sum_{N=0}^\infty N \frac{([I]\beta)^N}{N!} = \beta[I]~,
\end{equation}
where the probability density $[I]=-E_{\mathrm{inst}}$ of one instanton to form is given by Eq.~\eqref{eqn: kink energy} and
has to be assumed renormalised. As a consequence, the dilute gas approximation is valid if
\begin{equation}\label{eqn: diga condition}
    [I] \ll \omega~.
\end{equation}
The quantity $[I]/\omega$ is plotted in Figure~\ref{fig: diga validity}. We must compare this with the plot in the top panel of Figure~\ref{fig: T00 casimir vs tunnelling}. The effect of the instantons becomes relevant for $\lambda=1$ and $mL\sim1$. For these values of the parameters, the instanton rate $[I]$ takes values~$\sim 0.3\omega$, meaning that the dilute instanton gas approximation is decently good. 

\section{Details on the calculation of the Casimir energy}\label{app: casimir energy technical details}
In this section, we go through the details of the calculation of the Casimir energy, which has only been outlined in Section~\ref{sec: casimir energy}.

\subsection{$D=1$}
Let us begin by computing the Casimir energy on a circle. This has already been calculated in the past, however we perform the necessary steps here to make our calculation in higher dimensions more transparent.
We start by writing the vacuum energy on the circle
\begin{equation}
    E_0^{(1)} (m; L) = \sum_{n=-\infty}^\infty \frac{1}{2} \omega^{(1)}(m; n) \,,
\end{equation}
where $\omega^{(1)} (m; n)$ is the dispersion relation 
\begin{equation}
    \omega^{(1)} (m; n) = \sqrt{\left(\frac{2\pi n}{L}\right)^2 + m^2} 
\end{equation}
and $m$ is the mass.
To use the Abel-Plana formula~\eqref{eqn: Abel-Plana formula}, we must have a sum that goes from~0 to~$\infty$. The dispersion relation is symmetric, namely $\omega^{(1)}(m; n) = \omega^{(1)}(m; -n)$, thus we write
\begin{align}
    &\frac{1}{2} \sum_{n=-\infty}^\infty \omega^{(1)}(m; n) =  \frac{1}{2} \sum_{n=0}^\infty (2-\delta_{n,0}) \omega^{(1)}(m;n)\notag \\
    &=  \frac{1}{2} \left( 2 \int_0^\infty dt\, \omega^{(1)} (m;t) + \omega^{(1)}(m;0) + 2 i \int_0^\infty dt\, \frac{\omega^{(1)}(m;it) - \omega^{(1)}(m;-it)}{e^{2\pi t} - 1}  - \omega^{(1)}(m;0) \right)\notag \\
    &=  \frac{1}{2} \int_{-\infty}^\infty dt\, \omega^{(1)} (m;t) + i \int_0^\infty dt \, \mathrm{Disc}_+ \left[\frac{\omega^{(1)}(m;it)}{e^{2\pi t} - 1} \right]\,,
\end{align}
where in the last step we used that the function $\omega^{(1)}(m;it)$ has a branch cut on~$[t_+,+\infty)$, where~$t_+=\frac{mL}{2\pi}$.\footnote{To see this, first recall that when $f(t)$ has branch cuts on the imaginary axis, one can replace $i t$ and $-i t$ in the last term of the Abel-Plana formula with $i t+\epsilon$, $-it +\epsilon$ (with $\epsilon$ being a positive infinitesimal) respectively. Second, it is useful to write $\omega^{(1)}(m;t)=e^{1/2*\ln(m^2+t^2)}$. Then one can show that ${\omega^{(1)}(m;it+\epsilon)-\omega^{(1)}(m;-i t+\epsilon)=2i \sqrt{t^2-m^2}\theta(t-t_+)\equiv {\rm Disc}_+[\omega^{(1)}(m;it)]}$ using the analytic behavior of complex logarithm.}

A few comments are in place. First, we observe that there is no boundary term left, namely a term independent of $L$. This is because we are computing the energy on a circle, which has no boundary at all. This is a feature that will hold true in higher-dimensional cases as well.
Second, note how the divergent integral is exactly the continuous limit of the original sum. In particular
\begin{equation}
    \frac{1}{2} \int_{-\infty}^\infty dt\, \omega^{(1)} (m;t) = \: \frac{L}{2} \int_{-\infty}^\infty \frac{d k}{2\pi} \sqrt{k^2 + m^2} = \: E_\infty^{(1)} (m;L) \,,
\end{equation}
which is the vacuum energy on a circle of size $L$ when computed in the continuum limit.

We are left with computing the integral over the branch cut. We have
\begin{align}
    i \int_0^\infty dt \, \mathrm{Disc}_+ \left[\frac{\omega^{(1)}(m;it)}{e^{2\pi t} - 1} \right] = & i \int_0^\infty dt \, 2i \frac{\sqrt{\left(\frac{2\pi t}{L}\right)^2 - m^2}}{e^{2\pi t} - 1} \Theta(t-t_+) \notag \\
    = & - \frac{1}{\pi L} \int_{mL}^\infty dy \, \frac{\sqrt{y^2 - (mL)^2}}{e^{y}-1} \notag \\
    = & - \frac{1}{\pi L} F_{(1,0)}(m L) \,,
\end{align}
where we have defined the function 
\begin{equation}
    F_{(1,0)}(x) = \int_x^\infty dy \, \frac{\sqrt{y^2 - x^2}}{e^{y}-1}\,.
\end{equation}
The function above can also be written in terms of modified Bessel functions of the second type \cite{Bordag}
\begin{equation}
    F_{(1,0)}(x) = x \sum_{n=1}^\infty \frac{K_1(nx)}{n}\,.
\end{equation}

We can write the vacuum energy for the massive scalar field on a circle 
\begin{equation}
    E_0^{(1)} (m;L) = E_\infty^{(1)}(m;L) - \frac{1}{\pi L} F_{(1,0)}(m L)\,,
\end{equation}
and for the Casimir energy we obtain
\begin{equation}
    E_{\mathrm{Casimir}}^{(1)} (m;L) = - \frac{1}{\pi L} F_{(1,0)}(m L)\,.
\end{equation}

\subsection{Arbitrary number of dimensions: a recursive formula}

When we go higher in the number of dimensions we have more and more sums, and the procedure which we just employed above becomes increasingly cumbersome. Fortunately, we are able to give a recursive formula that will greatly simplify our calculations.

\emph{Ansatz:} Let $D>1$ be the number of spatial dimensions. Then, the Casimir energy for the real massive scalar field on the $D$-torus of sides $\{a_1,\ldots,a_D\}$ is given by
\begin{align}
    E_{\mathrm{Casimir}}^{(D)} (m; a_1, \ldots,a_D ) = \: & - \frac{1}{\pi^D a_D} \left(\prod_{i=1}^{D-1} \frac{a_i}{a_D}\right) F_{(D,0)}(ma_D) \notag \\
    & + \sum_{n=-\infty}^\infty E_{\mathrm{Casimir}}^{(D-1)} \left( \sqrt{m^2 + \frac{4\pi^2n^2}{a_D^2}}; a_1, \ldots,a_{D-1} \right) \,,
\end{align}
where
\begin{equation}
    F_{(D,0)} (x) = \: \left( \prod_{i=1}^{D-1} \int_0^\infty d y_i \right)  F_{(1,0)} \left( \scriptstyle{\sqrt{x^2 + \sum_{i=1}^{D-1}y_i^2} }\right) = \: \int_0^\infty d y F_{(D-1,0)} \left( \sqrt{x^2 + y^2} \right)\,.
\end{equation}
We prove the formula~\eqref{eqn: Casimir energy for generic D} by construction.
Let us assume we know the Casimir energy on the $D-1$-dimensional torus of sides $\{a_1,\ldots,a_{D-1}\}$. The dispersion relation is
\begin{equation}
    \omega^{(D-1)}(m;n_1,\ldots,n_{D-1} ) = \: \sqrt{\sum_{i=1}^{D-1} \left(\frac{2\pi n_i}{a_i}\right)^2 + m^2}\,.
\end{equation}
We then compute the Casimir energy on the $D$-torus of sides $\{a_1,\ldots,a_D\}$. The dispersion relation is
\begin{align}
    \omega^{(D)}(m;n_1,\ldots,n_{D} ) = \: & \sqrt{\sum_{i=1}^{D} \left(\frac{2\pi n_i}{a_i}\right)^2 + m^2} \notag \\
    = \: & \omega^{(D-1)}\left( \sqrt{m^2 + \frac{4\pi^2 n_D^2}{a_D^2}};n_1,\ldots,n_{D-1} \right)\,.
\end{align}
Using this, we can write the vacuum energy
\begin{align}
    E_0^{(D)} (m; a_1, \ldots,a_{D} ) = \: & \sum_{n_D=-\infty}^\infty \left( \sum_{n_1,\ldots,n_{D-1}=-\infty}^\infty \frac{1}{2} \omega^{(D)}(m;n_1,\ldots,n_{D-1},n_D) \right) \notag \\[2ex]
    = \: & \sum_{n_D=-\infty}^\infty E_0^{(D-1)} \left(\sqrt{m^2 + \frac{4\pi^2n_D^2}{a_{D}^2}}; a_1, \ldots,a_{D-1} \right) \notag \\
    = \: & \sum_{n_D=-\infty}^\infty \Bigg[ E_\infty^{(D-1)} \left(\sqrt{m^2 + \frac{4\pi^2n_D^2}{a_{D}^2}}; a_1, \ldots,a_{D-1} \right) \notag \\
    & + E_{\mathrm{Casimir}}^{(D-1)} \left(\sqrt{m^2 + \frac{4\pi^2n_D^2}{a_{D}^2}}; a_1, \ldots,a_{D-1} \right) \Bigg]\,.
\end{align}
The last line above already reproduces correctly the second term in our ansatz. We are left with evaluating the first term.
\begin{align}
    &\sum_{n_D=-\infty}^\infty E_\infty^{(D-1)} \Bigg(\sqrt{m^2 + \frac{4\pi^2n_D^2}{a_{D}^2}}; a_1, \ldots,a_{D-1} \Bigg) \notag \\
    = \: & \left( \prod_{i=1}^{D-1} a_i \int_{-\infty}^\infty \frac{d k_i}{2\pi} \right) \sum_{n_D=-\infty}^\infty \frac{1}{2} \sqrt{\frac{4\pi^2n_D^2}{a_{D}^2} + m^2 + \sum_{i=1}^{D-1} k_i^2} \notag \\
    = \: & \left( \prod_{i=1}^{D-1} a_i \int_{-\infty}^\infty \frac{d k_i}{2\pi} \right) E_0^{(1)} \left( \sqrt{m^2 + \sum_{i=1}^{D-1} k_i^2}; a_{D}\right) \notag \\
    = \: & \left( \prod_{i=1}^{D-1} a_i \int_{-\infty}^\infty \frac{d k_i}{2\pi} \right) \Bigg[ E_\infty^{(1)} \left( \sqrt{m^2 + \sum_{i=1}^{D-1} k_i^2}; a_{D} \right) + E_{\mathrm{Casimir}}^{(1)} \left( \sqrt{m^2 + \sum_{i=1}^{D-1} k_i^2}; a_{D} \right) \Bigg] \notag \\
    = \: & E_\infty^{(D)}(m;a_1,\ldots,a_{D}) + \left( \prod_{i=1}^{D-1} a_i \int_{-\infty}^\infty \frac{d k_i}{2\pi} \right) \left(-\frac{1}{\pi a_{D}}\right) F_{(1,0)} \left( a_{D} \sqrt{m^2 + \sum_{i=1}^D k_i^2} \right) \notag \\
    = \: & E_\infty^{(D)}(m;a_1,\ldots,a_{D}) - \frac{1}{\pi^{D} a_{D}} \left( \prod_{i=1}^{D-1} \frac{a_i}{a_{D}} \right) F_{(D,0)} (m a_{D})\,,
\end{align}
which concludes the proof.
Note that as $E_{\mathrm{Casimir}}^{(1)}(m;L) \propto e^{- m L}$ as $mL \gg 1$, it can be easily argued that the integral and the series appearing in formula~\eqref{eqn: Casimir energy for generic D} are convergent for any $D$.

\subsubsection{$D=2$}
We can use the formula~\eqref{eqn: Casimir energy for generic D} to obtain the Casimir energy for a massive scalar field on a two-torus. For the sake of generality, let us take a torus of linear sizes $a_1=a$ and $a_2=b$.
The formula then reads
\begin{align}
    E_{\mathrm{Casimir}}^{(2)} (m; a, b ) = \: & - \frac{1}{\pi^2 b} \frac{a}{b} F_{(2,0)}(mb) + \sum_{n=-\infty}^\infty E_{\mathrm{Casimir}}^{(1)} \left( \sqrt{m^2 + \frac{4\pi^2n^2}{b^2}}; a \right) \notag \\
    = & \: - \frac{1}{\pi^2 b} \frac{a}{b} F_{(2,0)}(mb) -\frac{1}{\pi a} \sum_{n=-\infty}^\infty F_{(1,0)}(\frac{a}{b} \sqrt{m^2b^2 + 4\pi^2n^2}) \notag \\
    = & \: - \frac{1}{\pi^2 a} \left[ \frac{a^2}{b^2} F_{(2,0)}(mb) + \pi F_{(1,1)} \left( \frac{a}{b} , mb \right) \right]\,.
\end{align}
Although not obvious, our result is indeed symmetric under exchanging $a$ and $b$.
We have defined the function
\begin{equation}
    F_{(1,1)} (r,x) = \: \sum_{n=-\infty}^\infty F_{(1,0)} (r \sqrt{x^2 + 4\pi^2n^2} )\,.
\end{equation}

\subsubsection{$D=3$}
Having derived the result for $D=2$, we can finally give a result for the Casimir energy for~$D=3$.
Again, we compute the Casimir energy of a massive scalar field on a three-torus with sizes $a_1=a$, $a_2=b$ and $a_3=c$.
Applying the formula~\eqref{eqn: Casimir energy for generic D}, we get
\begin{align}\label{eqn: casimir energy on a generic 3-torus}
    E_{\mathrm{Casimir}}^{(3)} (m; a, b , c) = \: & - \frac{1}{\pi^3 c} \frac{a}{c} \frac{b}{c} F_{(3,0)}(mc) + \sum_{n=-\infty}^\infty E_{\mathrm{Casimir}}^{(2)} \left( \sqrt{m^2 + \frac{4\pi^2n^2}{c^2}}; a , b \right) \notag \\
    = \: & - \frac{1}{\pi^3 c} \frac{a}{c} \frac{b}{c} F_{(3,0)}(mc) + \sum_{n=-\infty}^\infty \left(- \frac{1}{\pi^2 a}\right) \Bigg[ \frac{a^2}{b^2} F_{(2,0)}(\frac{b}{c} \sqrt{m^2c^2 + 4\pi^2n^2}) \notag \\
    & + \pi F_{(1,1)} \left( \frac{a}{b} ,\frac{b}{c} \sqrt{m^2c^2 + 4\pi^2n^2} \right) \Bigg] \notag \\
    = \: & - \frac{1}{\pi^3 a} \Bigg[ \frac{a^2}{c^2} \frac{b}{c} F_{(3,0)}(mc) + \pi \frac{a^2}{b^2} F_{(2,1)} \left( \frac{b}{c}, mc \right) + \pi^2 F_{(1,2)} \left( \frac{a}{b} , \frac{b}{c}, mc \right) \Bigg]\,.
\end{align}
We have introduced two new functions
\begin{align}
    F_{(2,1)}(r,x) = \: & \sum_{n=-\infty}^\infty F_{(2,0)} (r \sqrt{x^2+4\pi^2n^2}) \,, \\
    F_{(1,2)}(r_1,r_2,x) = \: & \sum_{n=-\infty}^\infty F_{(1,1)} (r_1, r_2 \sqrt{x^2+4\pi^2n^2}) \,.
\end{align}

\subsection{Comparing with the numerical renormalisation}
When computing the renormalised ratio of functional determinants appearing in the instanton contributions, we cannot use the Abel-Plana formula to obtain an analytic result. Instead, we will subtract the ultraviolet (UV) divergences via counter-terms, which amounts to subtracting the continuum limit from the series.
Therefore, it makes sense to compare this numerical procedure to the result we obtained for the Casimir energy, given that we only have analytical control for the latter. 
In order to do so, define the numerically renormalised Casimir energy as follows
\begin{equation}
    E_{\mathrm{Casimir}}^{(3)} (m; L) = \: \frac{1}{2} \sum_{n,\ell,\kappa = -\infty}^\infty \sqrt{\veck_{\{n,\ell,\kappa\}}^2 + m^2} e^{-\frac{\veck_{\{n,\ell,\kappa\}}^2}{\Lambda^2}} - \frac{L^3}{2} \int_{\mathbb{R}^3} \frac{d \veck^3}{(2\pi)^3} \sqrt{\veck^2 + m^2} e^{-\frac{\veck^2}{\Lambda^2}} \,.
\end{equation}
We have introduced the regularising function
\begin{equation}
    \eta (|\veck|) = e^{-\frac{\veck^2}{\Lambda^2}}\,,
\end{equation}
where $\Lambda$ is some large cut-off.
Also, we have denoted the discretised momentum by $\veck_{\{n,\ell,\kappa\}}^2 = \frac{4\pi^2}{L^2}(n^2+\ell^2+\kappa^2)$.
The continuum integral can be computed analytically and expanded for large $\Lambda/m$
\begin{align}
    &\frac{L^3}{2} \int_{\mathbb{R}^3} \frac{d^3 \veck}{(2\pi)^3} \sqrt{\veck^2 + m^2}  e^{-\frac{\veck^2}{\Lambda^2}} = \:  m \frac{(mL)^3}{16\pi^2} e^{\frac{m^2}{2\Lambda^2}} \frac{\Lambda^2}{m^2} K_1\left(\frac{m^2}{2\Lambda^2} \right)\notag \\[2ex]
    &\qquad\qquad\qquad\qquad= \:  m \frac{(mL)^3}{128\pi^2} \left[ \frac{16\Lambda^4}{m^4} + \frac{8\Lambda^2}{m^2} + 1 + 2\gamma_E + 2\log \frac{m^2}{4\Lambda^2} \right] + \mathcal{O}\left(\frac{m}{\Lambda}\right)\,.
\end{align}

\begin{figure}
    \centering
    \includegraphics[width=0.7\textwidth]{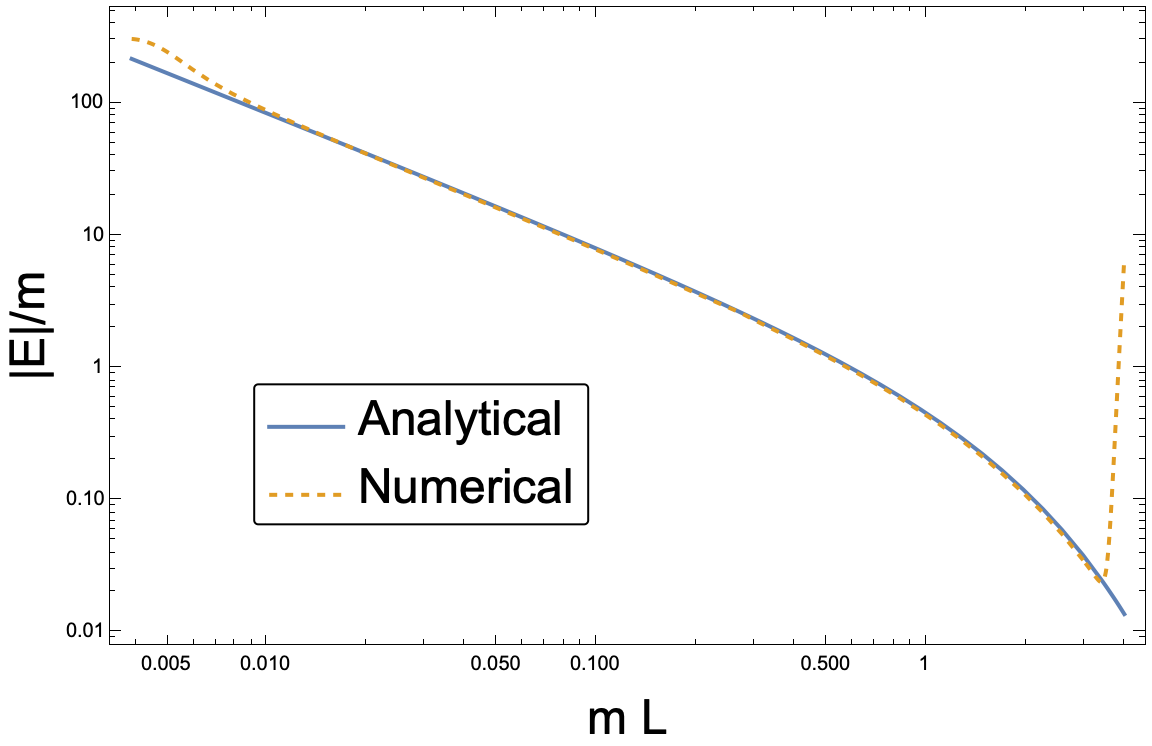}
    \caption{Comparison between the analytical and numerical renormalisation procedure for the Casimir energy.}
    \label{fig: analytical vs numerical casimir energy}
\end{figure}

As for the series, it is useful to introduce the degeneracy function $r_3(j)$, which returns the number of ways in which $j$ can be written as a sum of three squared integers. For example $r_3(0)=1$ and $r_3(1)=6$.
We can then write
\begin{equation}
    E_{\mathrm{Casimir}}^{(3)} (m; L) = \: \frac{1}{2} \sum_{j = 0}^{j_{\mathrm{max}}} r_3(j) \sqrt{\frac{4\pi^2}{L^2} j + m^2} e^{-\frac{4\pi^2 j}{L^2\Lambda^2}} + \mathrm{c.t.}\
\end{equation}
We introduced a cut-off $j_{\mathrm{max}}$ to the series. This has nothing to do with the regularisation cut-off $\Lambda$, and it is introduced to cut the numerical sum at a certain value. To make sure that it does not affect the renormalisation procedure, we must choose $j_{\mathrm{max}}$ such that
\begin{equation}
    \frac{4\pi^2}{L^2} j_{\mathrm{max}} \gg \Lambda^2 \gg \frac{4\pi^2}{L^2}\,,
\end{equation}
where the last inequality means that~$\Lambda$ must be large enough so that only the UV modes are affected by the regulator.
Then, we can compare the Casimir energy when obtained analytically versus numerically. This is shown in Figure~\ref{fig: analytical vs numerical casimir energy}.
The matching is pretty good but for the asymptotic regions $m L \ll 1$ and $m L \gg 1$. This is a feature of the numerical procedure.
To understand this, let us rewrite our conditions on the numerical cut-off~$j_{\mathrm{max}}$ and renormalisation cut-off~$\Lambda$ in a different way
\begin{equation}
    j_{\mathrm{max}} \gg \frac{m^2L^2}{4\pi^2} \frac{\Lambda^2}{m^2} \gg 1 \,.
\end{equation}
From the first of these inequalities, we see that as~$mL$ grows, we must increase~$j_{\mathrm{max}}$ more and more, making our numerical procedure more expensive. 
On the other hand, the second inequality tells us that what is a good renormalisation cut-off~$\Lambda$ for large~$mL$ becomes a bad cut-off for small $mL$.
Better agreement in the asymptotic regions can be achieved by splitting the numerical function into two regions and defining appropriate~$j_{\mathrm{max}}$ and~$\Lambda$ independently in either region. This is allowed since the result is cut-off independent.

\bibliographystyle{JHEP}
\bibliography{biblio.bib}

\end{document}